\documentclass[preprint,pra,showpacs,showkeys,floatfix]{revtex4}

\usepackage{graphicx}
\usepackage{dcolumn} 
\usepackage{array}
\usepackage{amsmath}
\usepackage{amsfonts}

\begin{document}

\title{Supersymmetry and eigensurface topology \\ of the planar quantum pendulum}

\author{Burkhard Schmidt}
\email{burkhard.schmidt@fu-berlin.de}
\affiliation{
Institute for Mathematics, Freie Universit\"{a}t Berlin \\ Arnimallee 6, D-14195 Berlin, Germany}

\author{Bretislav Friedrich}
\email{bretislav.friedrich@fhi-berlin.mpg.de}
\affiliation{
Fritz-Haber-Institut der Max-Planck-Gesellschaft \\ Faradayweg 4-6, D-14195 Berlin, Germany}

\date{\today}

\begin{abstract}

We make use of supersymmetric quantum mechanics (SUSY QM) to find three sets of conditions under which the problem of a planar quantum pendulum becomes analytically solvable. The analytic forms of the pendulum's eigenfuntions make it possible to find analytic expressions for observables of interest, such as the expectation values of the angular momentum squared and of the orientation and alignment cosines as well as of the eigenenergy. Furthermore, we find that the topology of the intersections of the pendulum's eigenenergy surfaces can be characterized by a single integer index whose values correspond to the sets of conditions under which the analytic solutions to the quantum pendulum problem exist.

\end{abstract}


\maketitle

\section{Introduction}

The quantum pendulum problem has a venerable history stretching back to the early days of quantum mechanics. First tackled by Condon in 1928 \cite{Condon:28a} in its planar variety, the quantum pendulum has since turned up in a number of research areas of atomic, molecular and optical physics, ranging from spectroscopy to the stereodynamics of molecular collisions to the manipulation of matter by external electric, magnetic and optical fields. Although both the planar and the full-fledged 3D spherical pendular varieties possess analytic asymptotic states \cite{Meixner:54a,McLachlan:64a,Abramovitz:72a,Gradshteyn:80a,Meyenn1970,SlenFriHerPRL1994,FriHerJPC95,FriHerZPhys96, HaerteltFriedrichJCP08}, the planar case has been explored with particular tenacity \cite{Nielsen:32a, Pradhan:73a,Cook:86a,Friedrich:91b,Baker:02a,Baker:05a,Dimeo:03a,Doncheski:03a,Leibscher:09a}, apparently because of its prototypical character, dwarfed only by that of few other systems such as of the harmonic oscillator.

The planar quantum pendulum or planar hindered rotor problem has been extensively employed  to model internal molecular rotation and molecular torsion in spectroscopy \cite{Herschbach1957, Herschbach1957b, Herschbach1959} and coherent control \cite{Seideman2007,Seideman2011, Stapelfeldt2009} as well as molecular orientation and alignment in spectroscopy and photodissociation dynamics \cite{HerFri1991,Friedrich:91b}. However, unlike the spherical pendulum, the planar pendulum has not been used so far to treat molecules subject to combined fields \cite{FriSlenHer1994, FriSlenHer1994b, SlenFriHer1994, FriHerJPCA99,FriNahBuck2003, PotFarBuckFri2008}. 

The spherical  quantum pendulum in combined fields has been the subject of a recent study based on supersymmetric quantum mechanics (SUSY QM)  \cite{s17,s26}, which resulted in finding an analytic solution to the problem for a particular class of states (the stretched states) and a particular ratio of the dimensionless parameters that characterize the strengths of the external fields that restrict the system's motion to to and fro pendular librations \cite{LemMusKaisFriPRA2011,LemMusKaisFriNJP2011}. A follow-up study \cite{SchmiFri2014a} revealed a close kinship between the conditions under which an analytic solution is obtained  and the topology of the intersections of the eigenenergy surfaces spanned by the dimensionless parameters. 

In this study, we seek -- and find -- analytic solutions for three sets of conditions that render the planar quantum pendulum problem (corresponding to a planar rotor subject to combined fields) analytically solvable and investigate the relationship between these sets of conditions and the topology of the planar pendulum's eigenenergy surfaces. We also make use of the analytic eigenfunctions to find the observables of interest -- such as the expectation values of the angular momentum squared and of the orientation and alignment cosines as well as of the eigenenergy -- likewise in analytic form. 

The Hamiltonian of the planar quantum pendulum problem has the general form
\begin{equation}
H = B\hat{J}^2 +V(\theta)
\label{eq:hamilton}
\end{equation}
where  $\hat{J}=-i\tfrac{d}{d\theta}$ is the angular momentum and  $B=\tfrac{\hbar^2}{2I}$ rotational constant with $I$ the moment of inertia. Note that in what follows we will assume $B=1$, which is equivalent to dividing all energies by $B$. In the Hamiltonian of Eq. (\ref{eq:hamilton}) the potential
\begin{equation}
V(\theta) = - \eta \cos \theta - \zeta \cos^2 \theta
\label{eq:pot}
\end{equation}
is restricted to the lowest two Fourier terms and $-\pi \le \theta \le \pi$ is a periodic coordinate. Since the $\cos\theta$ and $\cos^2\theta$ terms generate, respectively, oriented (single arrow-like) and aligned (double-arrow-like) states, we term the two interactions {\it orienting} and {\it aligning}. Their strengths are characterized, respectively, by the dimensionless parameters $\eta\geq 0$ and $\zeta\geq 0$. For  $\eta=\zeta=0$, the Hamiltonian of Eq. (\ref{eq:hamilton}) becomes that of a planar rotor.

This paper is organized as follows: In Sec. \ref{sec:Eigenproperties} we present, in turn, the cases of a purely orienting interaction, purely aligning interaction and of a combined orienting and aligning interaction and identify a condition for the intersections -- genuine or avoided -- of the eigenenergy surfaces. We find that the topology of the intersections can be characterized by a single topological index. In Sec. \ref {sec:susy}, we present three sets of conditions that lead to an analytic solution of the quantum pendulum problem and find that these conditions correspond to particular values of the topological index. In Sec. \ref{outlook}, we provide a summary of the present work and discuss its connections to related work. 

\section{Eigenproperties}
\label{sec:Eigenproperties}

\subsection{Pure orienting interaction:  $\eta > 0$ \& $\zeta=0$ }
\label{sec:eta}
When the planar rotor interacts with an external field solely via the orienting interaction, the Schr{\"o}dinger equation for Hamiltonian of Eq. (\ref{eq:hamilton}), $(H-E)\psi=0$, becomes isomorphic with the Mathieu equation \cite{Mathieu:68a}
\begin{equation}
	\frac{d^2 \psi}{dx^2}+[\lambda-2q\cos (2x)]\psi=0
	\label{eq:mathieu}
\end{equation}
whose characteristic values $\lambda$ for the interaction parameter $q=-2\eta$ and coordinate $x=\frac{1}{2}\theta$ are related to the eigenenergies $E$ by $\lambda=4E$, cf. the classic work on the quantum pendulum  \cite{Condon:28a,Nielsen:32a} and also ref. \cite{Friedrich:91b,Leibscher:09a}.
Then the required $2\pi$ periodicity, $\psi(\theta+2\pi)=\psi(\theta)$, of the problem in the original coordinate transforms into a $\pi$ periodicity, $\psi(x+\pi)=\psi(x)$, in the Mathieu coordinate.
Hence, the admissible solutions to the Schr{\"o}dinger equation for Hamiltonian (\ref{eq:hamilton}) with $\zeta=0$ are only the even-order Mathieu cosine elliptic, 
$ce_{2r}(\theta/2;-2\eta)$, or sine elliptic, $se_{2r+2}(\theta/2;-2\eta)$, functions.
Table \ref{tab:mathieu} lists the Fourier representations of these states (see the 1st and 4th rows) as well as the relationships between Mathieu functions with $q$ positive and negative (the latter needed here) that are obtained upon substituting $x\rightarrow(\pi/2-x)$ or $\theta\rightarrow(\pi-\theta$) \cite{Abramovitz:72a,Gradshteyn:80a,Gutierrez:03a}. 
In Table \ref{tab:mathieu} and in what follows the characteristic values $\lambda$ are referred to as $a$ and $b$ for the even ($ce$) and odd ($se$) parity eigenfunctions, respectively.

\begin{figure}
	\centering
		\includegraphics[width=12cm]{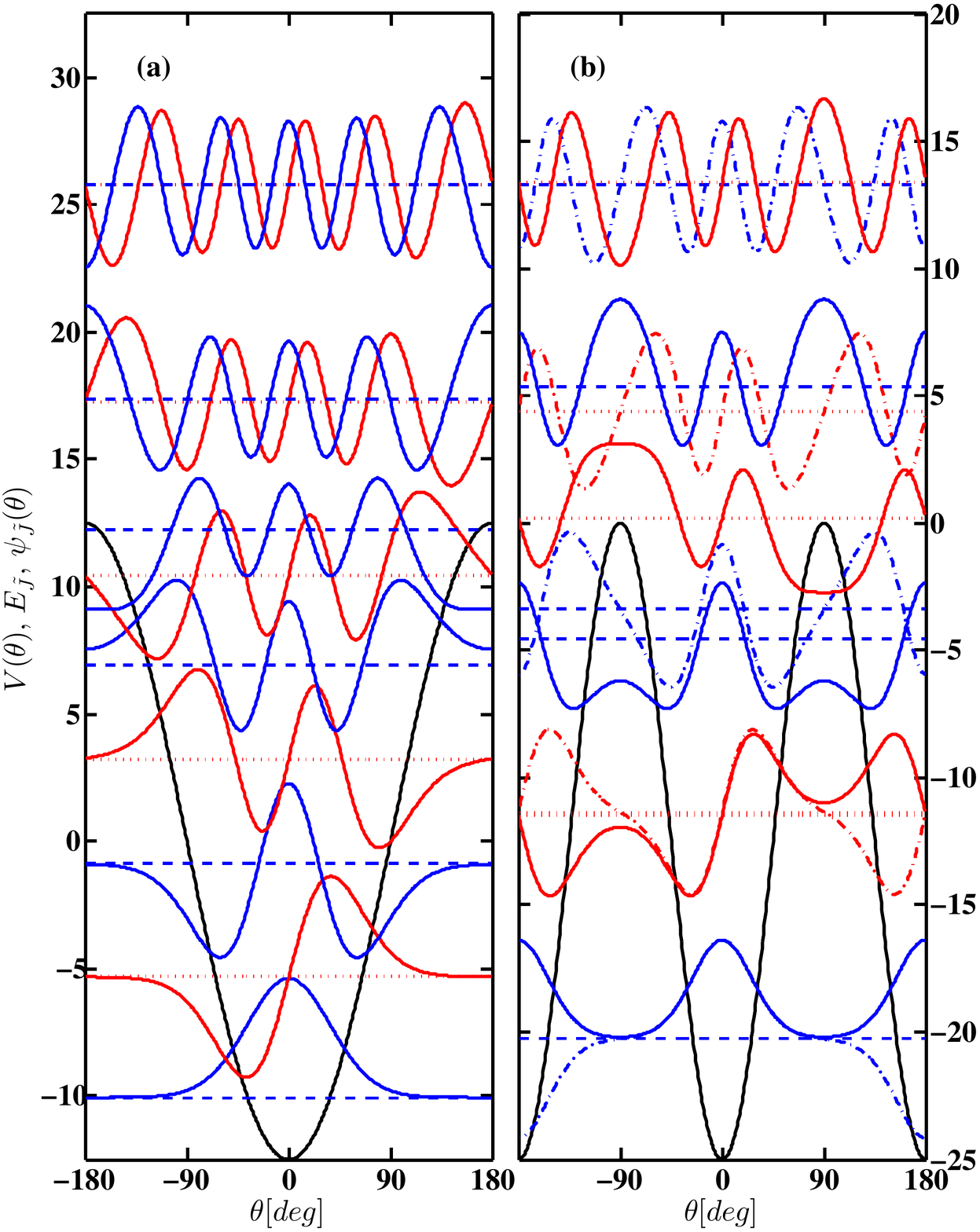}
	\caption{Eigenenergies and eigenfunctions pertaining to Hamiltonian  of Eq. (\ref{eq:hamilton}). 
	Left panel: purely orienting interaction with $\eta=12.5,\zeta=0$. 
	Right panel: purely aligning interaction with $\eta=0,\zeta=25$,
	where the tunneling doublets can be understood as zeroth-order intersection, $\kappa=0$, see Sec. \ref{sec:combined}. Wavefunctions of even and odd parity (with respect to $\theta=0$) are drawn in blue and red, respectively. In the right panel, full versus dash-dotted curves are used to distinguish between even and odd parity (with respect to $\theta=\pi/2$).}
	\label{fig:kappa0}
\end{figure}

The energy levels and wavefunctions for a pure orienting interaction are exemplified in the left panel of Fig. \ref{fig:kappa0} for $\eta=12.5$. While the states below the barrier qualitatively resemble those of a harmonic oscillator, the states above the barrier approach those of a free rotor, with (nearly) degenerate pairs of states of even and odd parity (with respect to $\theta=0$). 

\begin{figure}
  \centering
	\includegraphics[width=16cm]{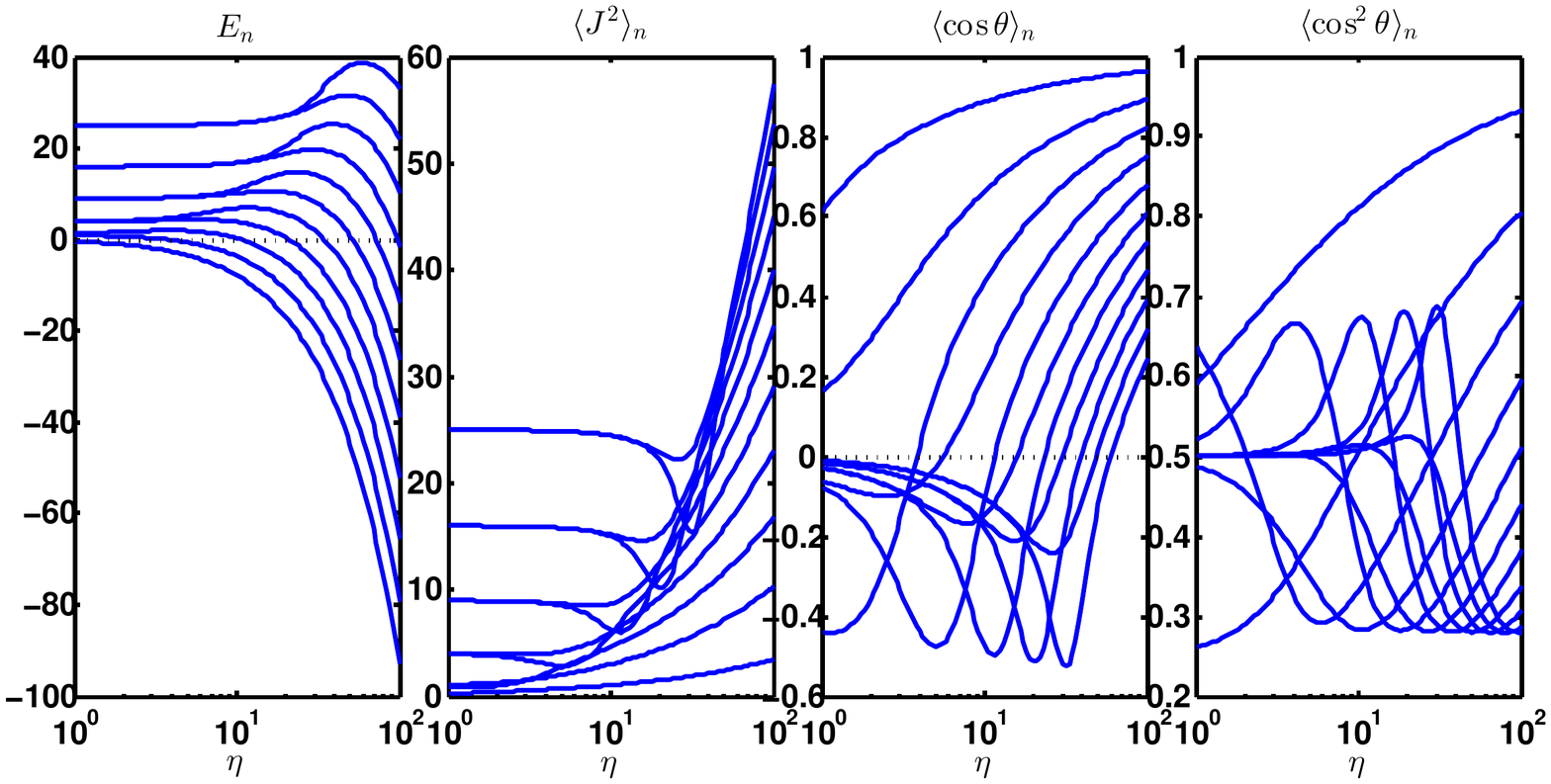}
	\caption{Eigenproperties of the planar quantum pendulum subject to a purely orienting interaction as a function of the interaction parameter $\eta$. The four panels show the eigenergies and the expectation values of the squared angular momentum, $\langle\text{J}^2\rangle$, orientation cosine, $\langle \cos\theta \rangle$, and alignment cosine, $\langle \cos^2\theta \rangle$.}
	\label{fig:eta}
\end{figure}

The dependence of the eigenenergies $E$ of the lowest eleven states on the orienting interaction parameter $\eta$ is shown in Fig. \ref{fig:eta}.
In the field-free limit, $\eta\rightarrow0$, the wavefunctions become those of a free planar rotor (i.e., ${ce}_r(x) \rightarrow \cos (rx)$ or ${se}_r(x)\rightarrow \sin (rx)$ for $r\neq 0$ and ${ce}_0(x)\rightarrow1/\sqrt{2}$); the corresponding energy levels approach a quadratic energy progression. In the strong field limit, $\eta\rightarrow \infty$, the eigenproperties become those of a harmonic angular oscillator (harmonic librator), which exhibits a linear progression of equidistant energy levels and a zero-point energy.
The asymptotic expansion of the characteristic values, Eq. (20.2.30) of ref. \cite{Abramovitz:72a}, yields the harmonic librator eigenenergies
\begin{equation}
	E_v \approx -\eta+\left(v+\frac{1}{2}\right)\sqrt{2\eta} 
\end{equation}
where $v=0,1,2,...$ is the harmonic librator quantum number and $\sqrt{2\eta}$ the librator quantum.

In Fig. \ref{fig:eta} we also show the expectation values of the angular momentum squared, $\langle \text{J}^2 \rangle$, and the directional characteristics of the states, the orientation cosine, $\langle \cos \theta \rangle$, and the alignment cosine, $\langle \cos^2\theta \rangle$, as functions of the orientation parameter $\eta$ for $\zeta=0$. Except for the lowest states of each symmetry (i.e., $ce_0$ and $se_0$), the states exhibit the Stern wrong-way orientation/alignment effect \cite{Friedrich:91b}: they become first anti-oriented, $\langle \cos\theta \rangle \rightarrow -1$, or anti-aligned, $\langle \cos^2\theta \rangle \rightarrow 0$, at low $\eta$, before conforming to the direction of the orienting field at large $\eta$. 

\subsection{Pure aligning interaction:  $\eta = 0$ \& $\zeta>0$ }
\label{sec:zeta}

\begin{figure}
  \centering
  \includegraphics[width=16cm]{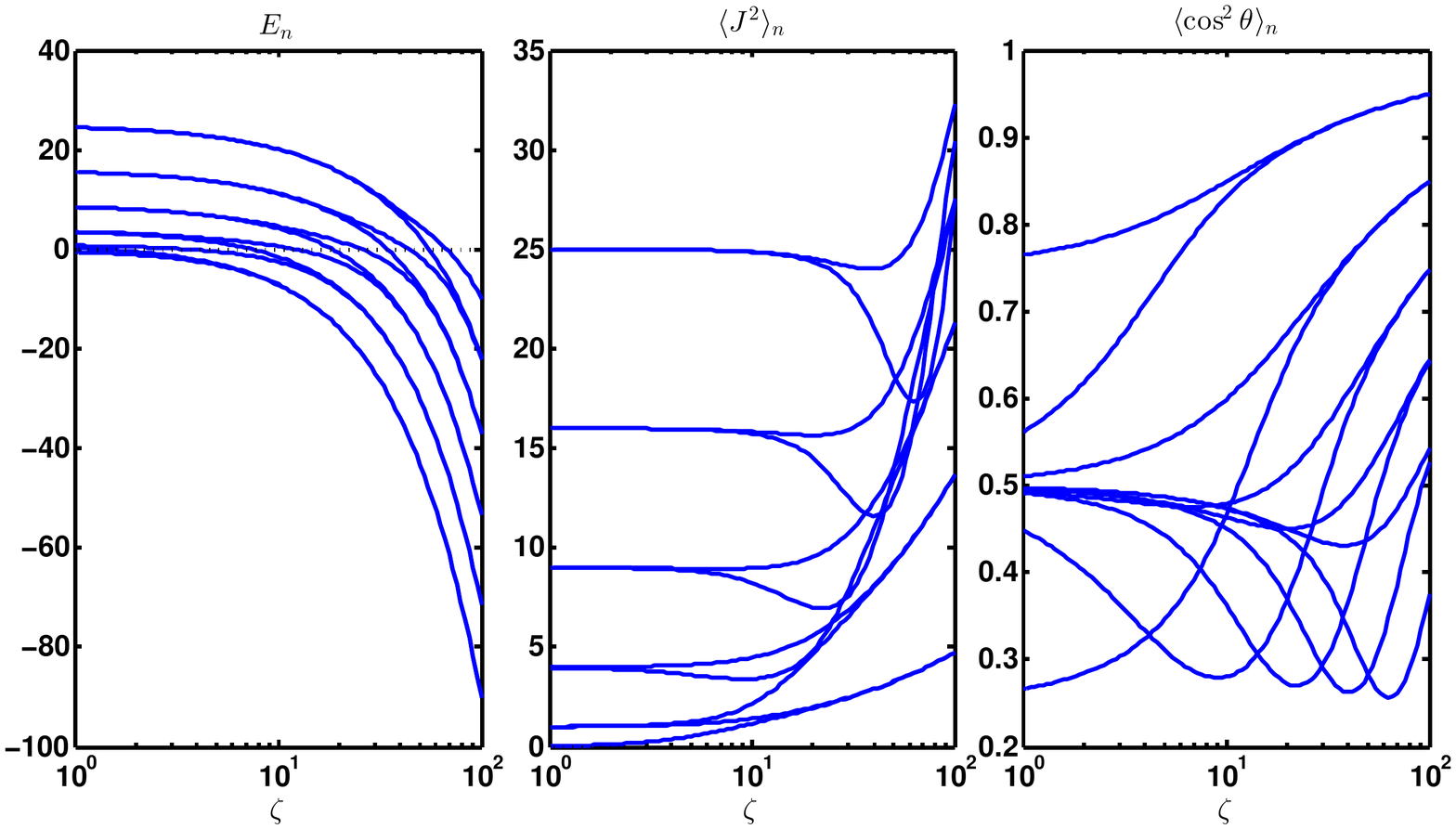}
	\caption{Eigenproperties of the planar quantum pendulum subject to a purely aligning interaction as a function of the interaction parameter $\zeta$. The three panels show the eigenenergies and the expectation values of the squared angular momentum $\langle\text{J}^2\rangle$ and alignment cosine $\langle \cos^2\theta \rangle$. Note that the orientation cosine vanishes identically.}
	\label{fig:zeta}
\end{figure}

Next, we consider the case when the planar rotor interacts with the external field solely via the aligning interaction, i.e., $\zeta>0$ and $\eta=0$. 
Also in this case, the corresponding Schr{\"o}dinger equation is isomorphic with the Mathieu equation (\ref{eq:mathieu})
but here the characteristic values $\lambda$ for a scaled interaction parameter $q=-\zeta/4$ and coordinate $x=\theta$ are related to the eigenenergies $E$ via $\lambda=E+\zeta/2$, see also ref. \cite{Friedrich:91b,Leibscher:09a}. 
Hence, the required $2\pi$ periodicity, $\psi(\theta+2\pi)=\psi(\theta)$, is satisfied for all four classes of Mathieu functions, i.e., for even- and odd-order Mathieu cosine-elliptic and sine-elliptic functions:  $ce_{2r}(\theta;-\zeta/4)$,  $se_{2r+1}(\theta;-\zeta/4)$, $ce_{2r+1}(\theta;-\zeta/4)$, and $se_{2r+2}(\theta;-\zeta/4)$, which are ordered here according to their increasing eigenenergy.  Table  \ref{tab:mathieu} lists these along with their symmetry and transformation properties for negative values of $q$. The corresponding eigenenergies as a function of the parameter $\zeta$ are displayed in Fig. \ref{fig:zeta}.

In the field-free limit, $\zeta \rightarrow0$, the eigenfunctions become trigonometric functions, see above,
with the eigenenergies forming a quadratic progression. 
In the harmonic librator limit, the asymptotic expansion of the characteristic values, Eq. (20.2.30) of ref. \cite{Abramovitz:72a}, yields an equidistant energy spectrum 
\begin{equation}
	E_{v}   \approx -\zeta+\left(v+\frac{1}{2}\right)\sqrt{4\zeta} 
\end{equation}
with a harmonic librator quantum $\sqrt{4\zeta}$ and quantum number $v=0,1,2,...$

The energy levels and wavefunctions for a purely aligning interaction ($\zeta>0$ and $\eta=0$) are exemplified in the right panel of Fig. \ref{fig:kappa0} for $\zeta=25$; their energies are also given in Tab. \ref{tab:ene}. Below the barrier, the levels $a_{2r}$ / $b_{2r+1}$ and $a_{2r+1}$ / $b_{2r+2}$ (for $r\ge0$) form pairwise near-degenerate tunneling doublets in the harmonic librator limit, split by tunneling through the equatorial barrier.  Hence the members of a given tunneling doublet correlate either with the even eigenfunctions $ce$ of odd order and the odd eigenfunctions $se$ of even order or with the even eigenfunctions $ce$ of even order and the odd eigenfunctions $se$ of odd order. 
The tunneling splitting in the harmonic librator limit, as obtained from 
Eq. (20.2.31) of ref. \cite{Abramovitz:72a}, becomes
\begin{equation}
	b_{2r+1}-a_{2r}=b_{2r}-a_{2r-1}\approx \frac{ 2^{3r+4} \zeta ^{\frac{r}{2}+\frac{3}{4}}e^{-2\zeta }}{\sqrt{\pi}r!}	
	\label{eq:mathieu_tunnel}
\end{equation}
The unbound states exhibit a free-rotor pairing of the levels $a_{2r}$ / $b_{2r}$ and $a_{2r+1}$ / $b_{2r+1}$ (for $r >0$) are pairwise degenerate in the field free limit. 
Right above the maximum of the potential, $V_{\text{max}}=0$, there is a single intermediate state at energy $E_6=0.1931$ (for the present example pertaining to $\zeta=25$), which is also listed in  Table \ref{tab:ene}.
We note that also the energy splittings of pairs of states well above the barrier converge to zero for increasing $\zeta$.

In Fig. \ref{fig:zeta} we also show the expectation values of the angular momentum squared, $\langle \text{J}^2 \rangle$, and the directional characteristic, the alignment cosine, $\langle \cos^2\theta \rangle$, as a function of the alignment parameter $\zeta$ for $\eta=0$. Again, all states except for the the lowest even and odd states ($ce_0$ and $se_1$) exhibit the Stern effect. Note that the orientation $\langle \cos \theta \rangle$ vanishes identically for $\eta=0$. 

\subsection{Combined orienting and aligning interactions:  $\eta > 0$ \& $\zeta>0$}
\label{sec:combined}

The potential of Hamiltonian (\ref{eq:hamilton}) exhibits two kinds of minima: 

\noindent (a) a global minimum for $\eta+2\zeta>0$ at  $\theta=0$ where the potential $V(\theta)$ can be approximated by
\begin{equation}
V(\theta) \approx -(\zeta+\eta)+\frac{1}{2}\left(2\zeta+\eta\right)\theta^2
\end{equation}
yielding bound states with energy levels
\begin{equation}
E_{v_0}=-(\zeta+\eta) + \left(v_0+\frac{1}{2}\right)\sqrt{4\zeta+2\eta}
\label{eq:E_harm_0}
\end{equation}
and vibrational quanta $\sqrt{4\zeta+2\eta}$. 

\noindent (b) a local minimum for $2\zeta>\eta$ at $\theta=\pi$ in whose vicinity the potential can be approximated by
\begin{equation}
V(\theta) \approx -(\zeta-\eta)+\frac{1}{2}\left(2\zeta-\eta\right)(\theta-\pi)^2
\end{equation}
yielding bound states with energy levels
\begin{equation}
E_{v_\pi}=-(\zeta-\eta) + \left(v_\pi+\frac{1}{2}\right)\sqrt{4\zeta-2\eta}
\label{eq:E_harm_pi}
\end{equation}
and vibrational quanta $\sqrt{4\zeta-2\eta}$. 

We note that the maximum, $V_{\text{max}}=\frac{\eta^2}{4\zeta}$, of the potential for the combined orienting and aligning interaction is located at $\theta= \arccos \left[-\frac{\eta}{2\zeta}\right]$; its position shifts with decreasing $\eta$ toward $\theta=\pi/2$ and with increasing $\eta$ toward $\theta=\pi$. However, in order for that to be the case, $\eta$ must not exceed $2\zeta$, as there would be neither a maximum nor a local minimum at $\theta=\pi$.

One can also use Eqs. (\ref{eq:E_harm_0}) and (\ref{eq:E_harm_pi}) to find a relationship between the interaction parameters $\eta$ and $\zeta$ at the loci of the intersection of the $E_{v_0}$ and $E_{v_\pi}$ energy levels. 
For $2\zeta\gg\eta$ the vibrational quanta become approximately equal, and the condition, $E_{v_0}=E_{v_\pi}$, for the degeneracy of levels localized around the global and local minima 
\begin{equation}
-(\zeta+\eta) + \left(v_0+\frac{1}{2}\right)\sqrt{4\zeta}=-(\zeta-\eta) + \left(v_\pi+\frac{1}{2}\right)\sqrt{4\zeta}
\end{equation}
yields
\begin{equation}
	\zeta = \left(\frac{\eta}{\kappa}\right)^2
	\label{eq:loci}
\end{equation}
with $\kappa$ the difference in the quantum numbers
\begin{equation}
\kappa=v_0-v_\pi
\label{eq:kappa}
\end{equation} 

\begin{figure}
  \centering
  \includegraphics[width=12cm]{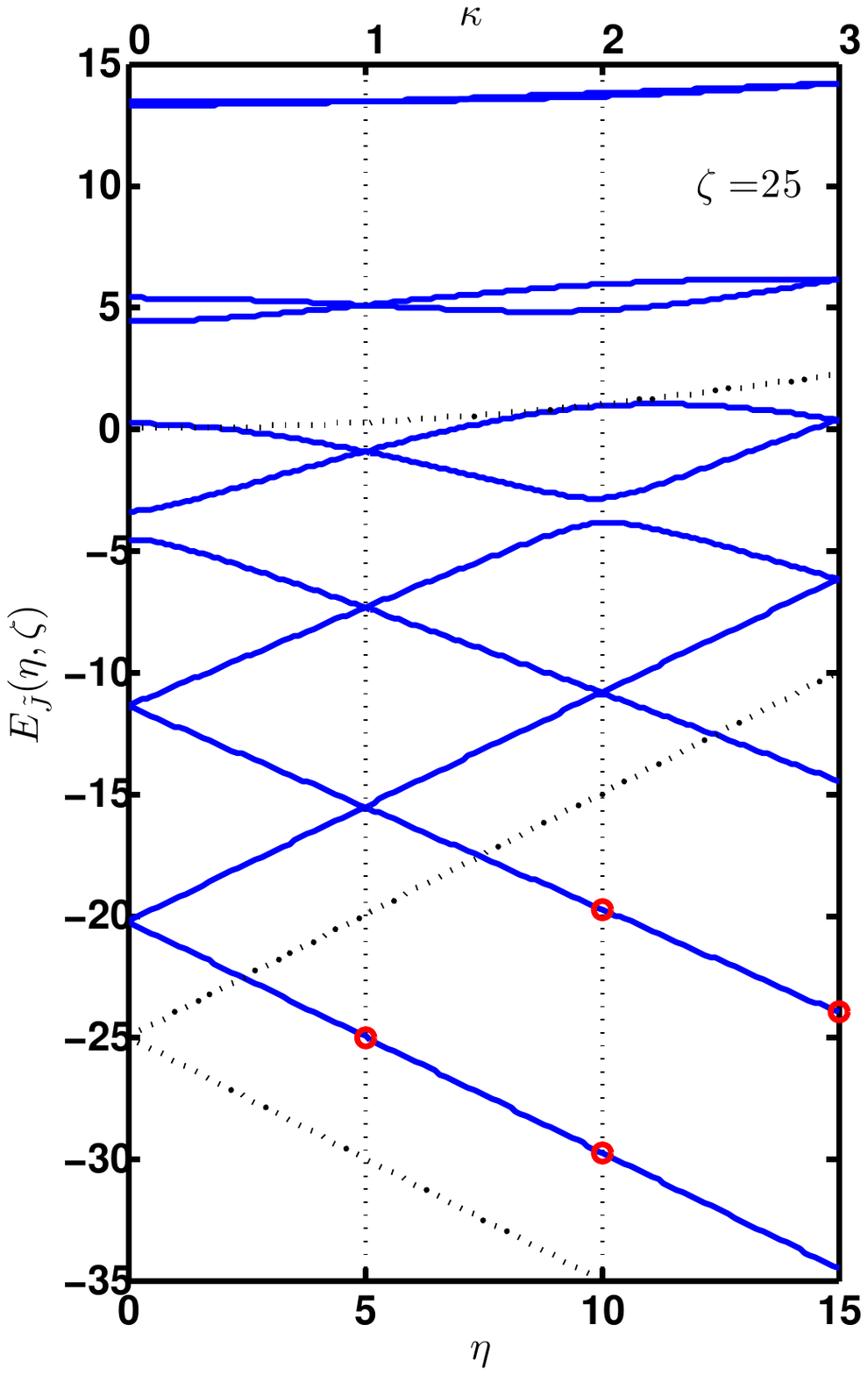}
  \caption{Eigenenergies of a planar rotor subject to the combined orienting and aligning interactions as a function of $\eta$ for a fixed $\zeta=25$. The dashed lines indicate (from bottom to top), respectively, the global minimum at $-(\zeta+\eta)$, the local/secondary minimum at $-(\zeta-\eta)$, and the maximum at $\frac{\eta^2}{4\zeta}$. Also shown are the values of the index $\kappa$ which defines the loci of the intersections. 
For $\kappa$ even, the intersections are avoided, for $\kappa$ odd they are genuine. The red circles indicate the four cases for which analytic solutions have been found via SUSY, see Sec. \ref{sec:susy}.}
  \label{fig:cut}
\end{figure}

\begin{figure}
  \centering
  \includegraphics[width=7cm]{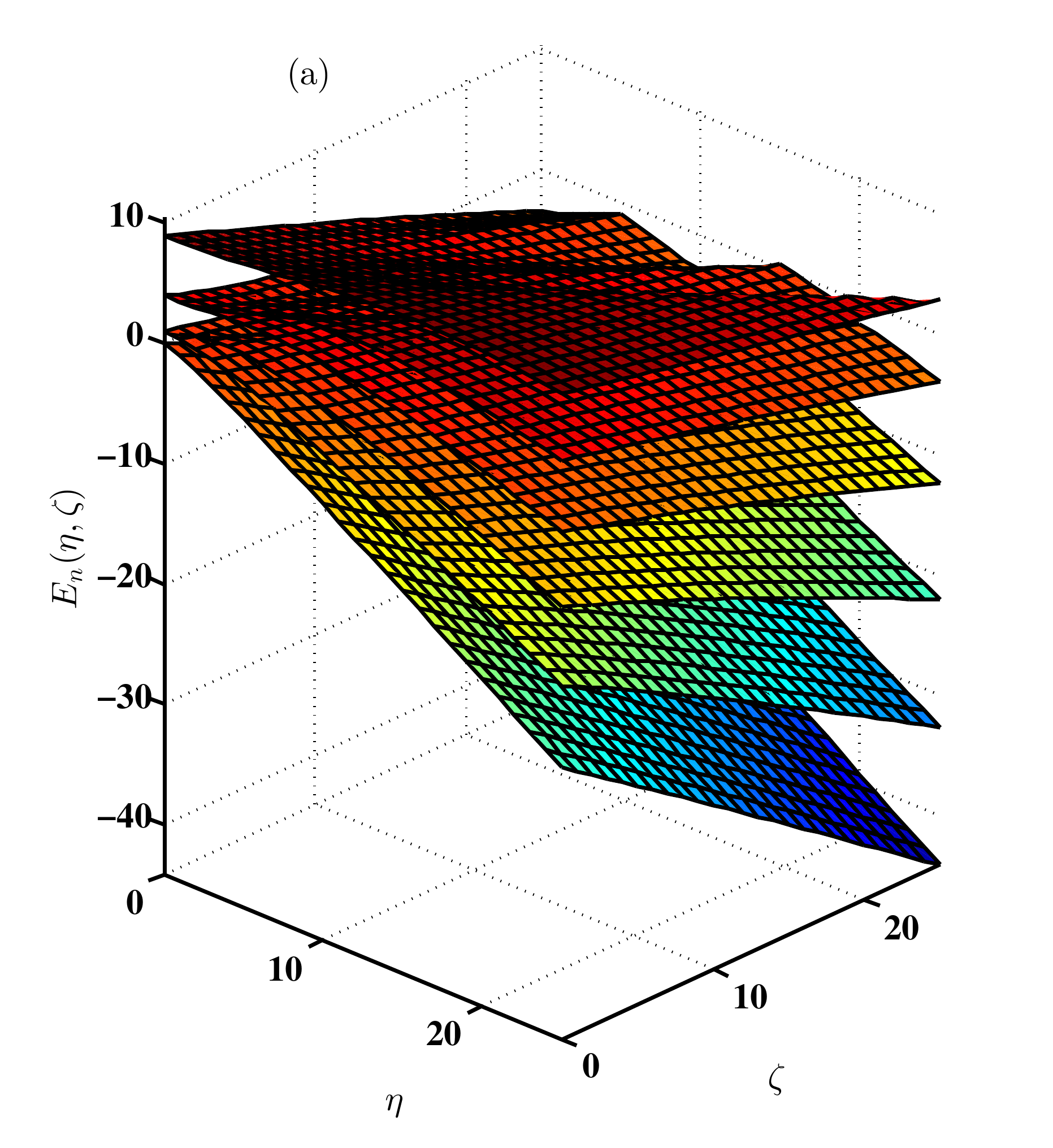}
  \includegraphics[width=7cm]{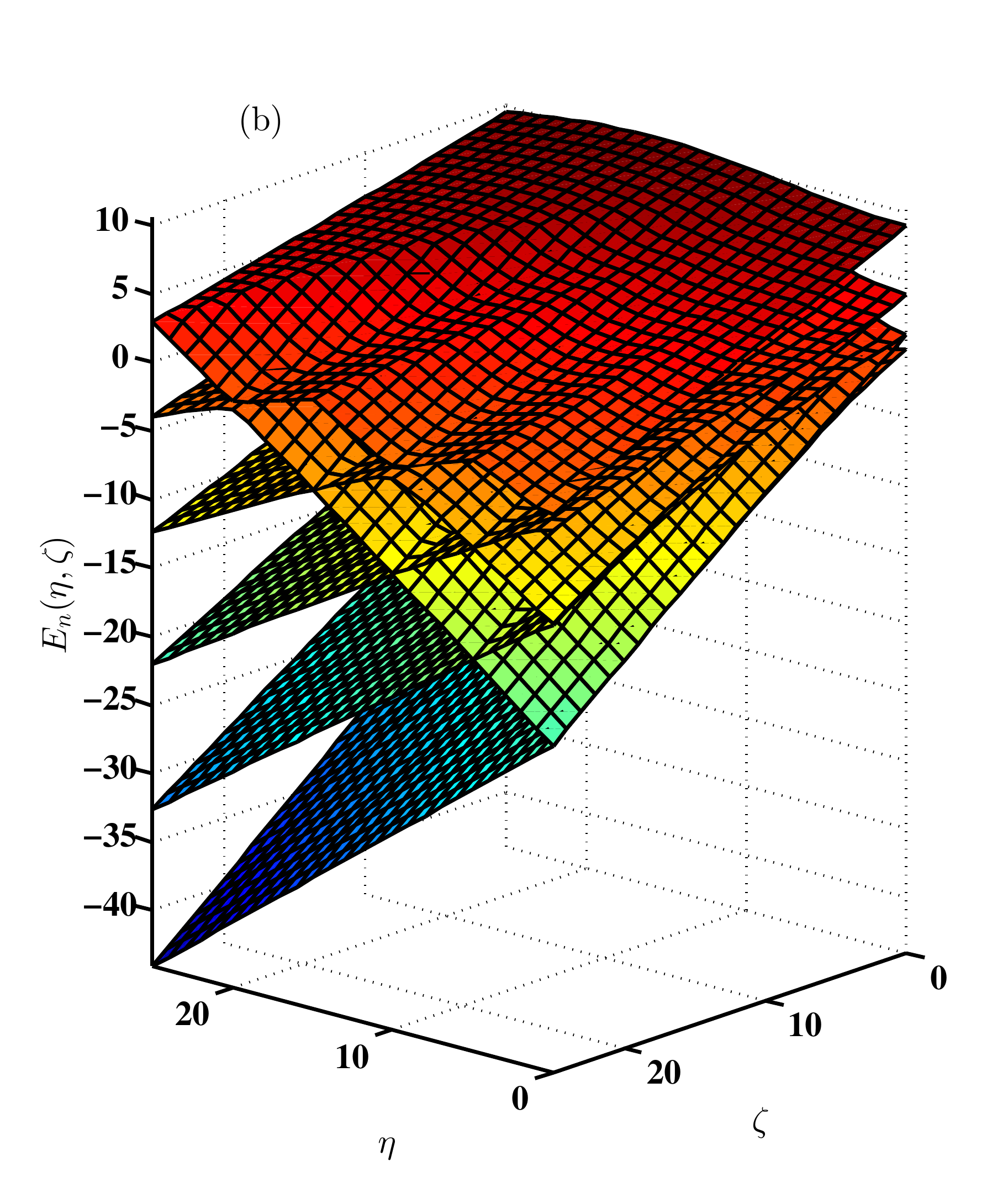}
  \caption{Views of the lowest six eigenenergy surfaces, $n=0,1,2,3,4,5$, of  Hamiltonian (\ref{eq:hamilton}) for a planar quantum pendulum  subject to combined orienting and aligning interactions. The  energy surfaces are shown as functions of the parameters $\eta$ and $\zeta$ that characterize the strengths of, respectively, the orienting and aligning interaction.}
  \label{fig:ene}
\end{figure}

The index $\kappa$ thus defines combinations of the interaction parameters $\eta$ and $\zeta$ at which the eigenenergy surfaces spanned by $\eta$ and $\zeta$ intersect. 
Fig. \ref{fig:cut} shows the eigenenergies of a planar rotor subject to the combined orienting and aligning interactions as a function of $\eta$ for a fixed $\zeta=25$. 
The dashed lines (from bottom to top) indicate, respectively, the global minimum at $-(\zeta+\eta)$, the local/secondary minimum at $-(\zeta-\eta)$, and the maximum at $\frac{\eta^2}{4\zeta}$ of potential (\ref{eq:pot}). Also shown are the values of the topological index $\kappa$: depending on whether $\kappa$ is even or odd, the corresponding intersections are found to be, respectively, avoided or genuine. 
This alternating pattern of avoided and genuine crossings follows from the symmetries of the intersecting states, as described in Sec. III. 
Fig. \ref{fig:ene} shows the eigenenergy surfaces spanned by the parameters $\eta$ and $\zeta$ pertaining to the lowest six eigenstates of a planar rotor subject to the combined interactions. 
As one can see, at $\zeta=0$ or $\eta=0$, the energy surfaces correspond to the Mathieu cases for the purely orienting or purely aligning interactions described above and shown in Figs. \ref{fig:eta} and \ref{fig:zeta}. 
For $\zeta\ge0$ and $\eta\ge0$, the eigenenergy surfaces exhibit the said intersections, which are found to occur exactly at the loci predicted by Eq. (\ref{eq:loci}) for a given $\kappa$. As Eq. (\ref{eq:kappa}) is state-independent, the number of intersections an energy surface partakes in is equal to the label $n$ of the corresponding eigenstate: 
the lowest energy surface, with $n=0$, is thus not involved in any intersection; 
the first excited state surface, with $n=1$, is involved in a first-order ($\kappa=1$) intersection (between nearest doublets); 
the second excited state surface, with $n=2$, is involved both in a first-order ($\kappa=1$) genuine intersection (between nearest doublets) and in a second-order ($\kappa=2$) avoided intersection (between second nearest doublets), etc. Consequently, at the loci of the $\kappa$-th order intersections given by Eq.~(\ref{eq:loci}), we find an energy level pattern with $\kappa$ single states at the bottom, followed by all other states which are doubly degenerate. In contrast, there are no degeneracies arising anywhere in between these intersection loci.

\begin{figure}
\centering
\includegraphics[width=12cm]{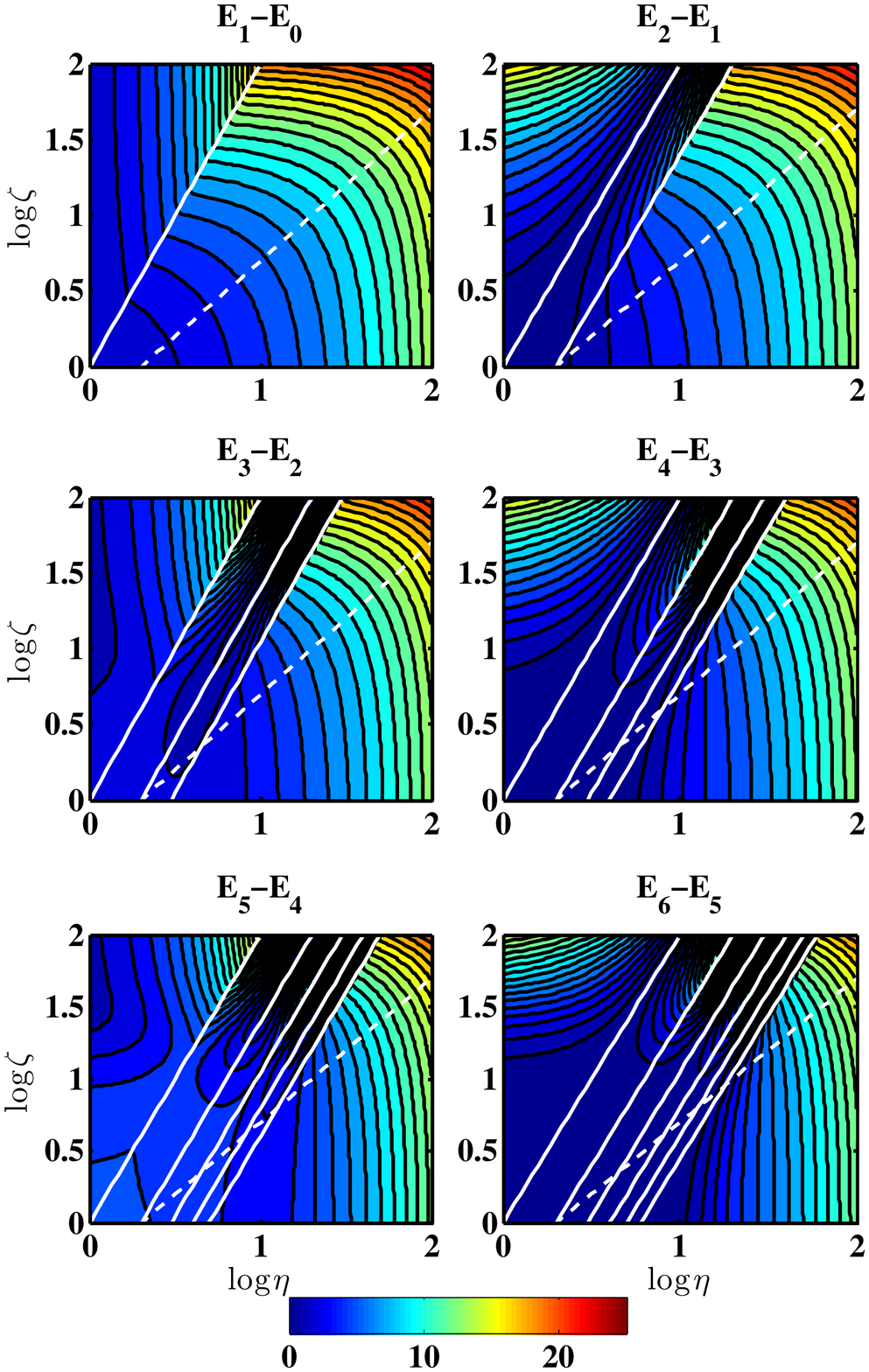}
\caption{Energy differences (gaps) between adjacent eigenenergy surfaces of Hamiltonian (\ref{eq:hamilton}) for a planar quantum pendulum  subject to combined fields. The  energy gaps are shown as functions of the parameters $\eta$ and $\zeta$ that characterize the strengths of, respectively, the orientation and alignment interactions. White lines indicate the loci of the $\kappa$-th order  intersection of adjacent surfaces, see Eq. (\ref{eq:loci}). The dashed line at
$\eta = 2\zeta$ marks the boundary above which the potential exhibits both a maximum and a local minimum, see Sec. \ref{sec:combined}.}
\label{fig:gap}
\end{figure}

The intersections of the eigenenergy surfaces are visualized in Fig. \ref{fig:gap} which shows the energy differences (gaps) between adjacent eigenenergy surfaces. The dashed line at
$\eta = 2\zeta$ marks the boundary of the condition $\eta \le 2\zeta$ at which the potential exhibits both a maximum, $V_{\text{max}} = \frac{\eta^2}{4\zeta}$, and a local minimum. By substituting the condition for the intersection loci, Eq. (\ref{eq:loci}), we obtain  
\begin{equation}
V_{\text{max}} = \frac{1}{4}\kappa^2
\label{eq:Vmax}
\end{equation}
which is independent of either of the interaction parameters but only reflects the way in which they combine at the intersections.
The full lines correspond to various values of the topological index $\kappa$ and thus of $V_{\text{max}}$.
Therefore, the dashed line crosses the $\kappa=1$ line where the energy of the lowest doublet (states 1 and 2) coincides with the maximum of the potential; only a single state (state $n=0$) lies below that energy.  
Analogously, the dashed line crosses the $\kappa=2$ line where the energy of the lowest doublet (states $n=2$ and $n=3$) coincides with $V_{\text{max}}$ (and two single states, 0 and 1, lie below that energy), etc. 

In the right panels of Fig. \ref{fig:gap}, the zero gap (darkest blue color) extends along the odd $\kappa$ lines all the way down to the field-free limit, whereas, in the left panels, the energy gaps along the even $\kappa$ lines increase slightly when approaching the field-free limit. 
Thereby the even and odd lines connect, respectively, to the genuine and avoided intersections of the planar rotor levels in the field-free limit, cf. Fig. \ref{fig:cut}. 
Note that the planar case differs in this respect from the spherical case where all intersections are avoided, cf. Fig. 3 of ref. \cite {SchmiFri2014a}.

\section{Supersymmetry of the planar quantum pendulum}
\label{sec:susy}

Supersymmetric quantum mechanics \cite{s17,s26} is based on the 
concept of superpartner Hamiltonians with corresponding Schr{\"o}dinger equations
\begin{eqnarray}
H_1 \psi_n^{(1)} &=& (A^\dagger A+\epsilon) \psi_n^{(1)} = E_n^{(1)} \psi_n^{(1)} \nonumber \\
H_2 \psi_n^{(2)} &=& (A A^\dagger+\epsilon) \psi_n^{(2)} = E_n^{(2)} \psi_n^{(2)}
\label{eq:TISE12}
\end{eqnarray}
where the symmetry of the construction ensures that  
\begin{eqnarray}
H_1 (A^\dagger \psi_n^{(2)}) &= (A^\dagger A A^\dagger+\epsilon A^\dagger)\psi_n^{(2)} =& E_n^{(2)} (A^\dagger \psi_n^{(2)} ) \nonumber \\
H_2 (A         \psi_n^{(1)}) &= (AA^\dagger A+\epsilon A)\psi_n^{(1)} =& E_n^{(1)} (A \psi_n^{(1)})
\label{eq:H1_H2}
\end{eqnarray}
which serves to establish relations between the eigenvalues, $E^{(1)},E^{(2)}$, and eigenfunctions, $\psi^{(1)},\psi^{(2)}$ of the superpartner Hamiltonians $H_1,H_2$, of Eq. (\ref{eq:TISE12}).
 For the usual choice of the constant $\epsilon$ being the $H_1$ ground state energy, $\epsilon=E_0^{(1)}$, this leads to
\begin{eqnarray}
E_n^{(2)}&=& E_{n+1}^{(1)} \nonumber \\
\psi_{n  }^{(2)} &\propto&  A         \psi_{n+1}^{(1)} \nonumber \\
\psi_{n+1}^{(1)} &\propto&  A^\dagger \psi_{n  }^{(2)}
\label{eq:twins}
\end{eqnarray}
i.~e., the SUSY partner Hamiltonians are isospectral, where the intertwining operators $A$ (or $A^\dagger$) convert the eigenfunctions of $H_1$ (or $H_2$) into those of $H_2$ (or $H_1$), at the same time lowering (or raising) the respective quantum numbers by one; only the ground state eigenfunction of $H_1$ lacks a partner state but is annihilated by the intertwining operator, $A\psi_0^{(1)}=0$. In contradistinction, other choices of $\epsilon$ give often rise to a partial or complete breakdown of the degeneracy of the energy levels of $H_1$ and $H_2$ \cite{s17,s26}, as encountered in some of the cases studied below.

For applications to SUSY QM in position representation, the standard choice of the intertwining operators is 
\begin{eqnarray}
A         &=&+\frac{d}{d\theta}+W(\theta) \nonumber \\
A^\dagger &=&-\frac{d}{d\theta}+W(\theta)
\label{eq:A}
\end{eqnarray}
which leads to the following expressions for superpartner Hamiltonians
\begin{eqnarray}
H_1&=&-\frac{d^2}{d\theta^2}+V_1(\theta) \nonumber \\
H_2&=&-\frac{d^2}{d\theta^2}+V_2(\theta)
\label{eq:H12}
\end{eqnarray}
where the supersymmetric partner potentials $V_1$ and $V_2$ are related to the superpotential $W(\theta)$ via Riccati-type equations
\begin{eqnarray}
V_1&=&W^2(\theta) - \frac{d}{d\theta}W(\theta) + \epsilon \nonumber \\
V_2&=&W^2(\theta) + \frac{d}{d\theta}W(\theta) + \epsilon
\label{eq:riccati}
\end{eqnarray}
For a nodeless ground state wavefunction, $\psi_0^{(1)}$, this allows to directly calculate the superpotential from a known ground state wavefunction
\begin{equation}
W(\theta)=-\frac{\frac{d}{d\theta}\psi_0^{(1)}}{\psi_0^{(1)}}
\end{equation}
which can be inverted to obtain an analytic expression for the wavefunction provided the superpotential is  known 
\begin{equation}
\psi_\epsilon^{(1)}(\theta) \propto \exp \left( -\int_0^\theta W(y)dy \right) 
\label{eq:psi}
\end{equation}
While this yields nonsingular superpotentials for the standard choice of the ground state, $\epsilon=E_0^{(1)}$, singularities of the superpotentials are encountered when choosing an excited state, $\epsilon=E_n^{(1)}$ with $n>0$, where the singularities arise at the zeros of the excited state wavefunctions. As a result, the partner Hamiltonians $H_1$ and $H_2$  are no longer isospectral \cite{s17,s26}, see also our results in Secs. \ref{sec:kappa2} and \ref{sec:kappa3}.

Throughout what follows we make use of the following {\it Ansatz} for the superpotential 
\begin{equation}
W(\theta) = \alpha\cot\theta+\beta\sin\theta+\gamma\csc\theta
\label{eq:W}
\end{equation}
which is an extension ($\gamma$ term added) of the {\it Ansatz} employed in refs. \cite{LemMusKaisFriPRA2011,LemMusKaisFriNJP2011} for the case of the spherical pendulum. Note that the $\alpha$ term alone is related to the Rosen-Morse I superpotential whereas a combination of the $\alpha$ and $\gamma$ terms bears similarity with the P\"{o}schl-Teller I superpotential, cf. ref. \cite{s17}.

Then Eq. (\ref{eq:W}) yields the following expressions for the SUSY partner potentials
\begin{eqnarray}
W^2(\theta) \mp W'(\theta) 
&=& (\alpha^2+\gamma^2\pm \alpha)\csc^2\theta \nonumber \\
&+& (2\alpha\gamma\pm \gamma)\cot\theta\csc\theta \nonumber \\
&-& (\pm \beta-2\alpha\beta)\cos\theta \nonumber \\
&-& \beta^2\cos^2\theta \nonumber \\
&-& (\alpha^2-\beta^2-2\beta\gamma)
\label{eq:V12}
\end{eqnarray}
By identifying the potential of Eq. (\ref{eq:pot}) for the quantum rotor subject to the combined interactions with $V_1=W^2-W'+\epsilon$, we obtain:
\begin{eqnarray}
\eta     &=& \beta-2\alpha\beta \nonumber \\
\zeta    &=& \beta^2 \nonumber \\
\epsilon &=& \alpha^2-\beta^2-2\beta\gamma
\label{eq:eze}
\end{eqnarray}
In order for the first two terms on the right hand side of Eq. (\ref{eq:V12}) to vanish, one of the following three conditions must be fulfilled: (A) $\alpha=0$ and $\gamma=0$; or (B) $\alpha=-1/2$ and $\gamma=\pm 1/2$; or (C) $\alpha=-1$ and $\gamma=0$. 
Below, we will discuss {\it cases A} through {\it C} in turn and show that each is connected with a particular ratio of $\eta$ to $\zeta$ and, therefore, with a particular topology of the eigenenergy surfaces, namely {\it case A} with $\kappa=1$, {\it case B} with $\kappa=2$, and {\it case C} with $\kappa=3$. 

The knowledge of the superpotential $W$ makes it possible to construct the supersymmetric partner potential $V_2=W^2+ W'+\epsilon$, which -- apart from the singular terms proportional to $\csc \theta$ -- differs from $V_1$ in that the interaction parameter $\eta$ is effectively reduced by $2\beta$. This makes the partner potential $V_2$ less asymmetric than the original potential $V_1$.

Furthermore, using Eq. (\ref{eq:psi}) one can derive an analytic expression   
for the wavefunction  from the superpotential $W$ pertaining to the energy eigenvalue $\epsilon$ as obtained from Eq. (\ref{eq:eze}).
For the particular superpotential introduced by Eq. (\ref{eq:W}), the wavefunction takes the form
\begin{eqnarray}
\psi_\epsilon(\theta) 
&\propto& (\csc\theta)^\alpha\,\exp(\beta\cos\theta)\,\left(\cot\frac{\theta}{2}\right)^\gamma\nonumber \\ 
&\propto& \left(\csc\frac{\theta}{2}\right)^{\alpha+\gamma}\,\left(\sec\frac{\theta}{2}\right)^{\alpha-\gamma}\,\exp(\beta\cos\theta)
\label{eq:psi_abc}
\end{eqnarray}

We note that identifying  potential  (\ref{eq:pot}) with $V_2=W^2+W'+\epsilon$ furnishes no new superpotentials and thus no new analytic wavefunctions. 

\subsection{First-order intersections or {\boldmath $\kappa=1$}}
\label{sec:kappa1}

For $\alpha=\gamma=0$ the superpotential (\ref{eq:W}) simplifies to
\begin{equation}
W=\beta\sin\theta
\label{eq:W_1}
\end{equation}
and Eq. (\ref{eq:eze}) yields the following expressions for the interaction parameters and the energy in terms of the parameter $\beta$ of Eq. (\ref{eq:W}),
\begin{eqnarray}
\eta&=&\beta \nonumber \\
\zeta&=&\beta^2 \nonumber \\
\epsilon &=& -\beta^2
\label{eq:eze_1}
\end{eqnarray}
In this case ({\it case A}), Eq. (\ref{eq:psi_abc}) yields an eigenfunction of the original Hamiltonian $H_1$
\begin{equation}
\psi_\epsilon^{(1)}(\theta) \propto \exp (\beta\cos\theta)
\label{eq:psi_1}
\end{equation}
which exhibits a pronounced maximum at $\theta=0$, i.~e., at the global minimum of the potential, and decays quickly for larger values of the angle $\theta$. Since $\psi_\epsilon^{(1)}$ is nodeless, we conclude that it corresponds to the ground state wavefunction $\psi_\epsilon^{(1)}=\psi_0^{(1)}$ pertaining to the ground-state energy $\epsilon=E_0^{(1)}$.

\begin{figure}
	\centering
		\includegraphics[width=12cm]{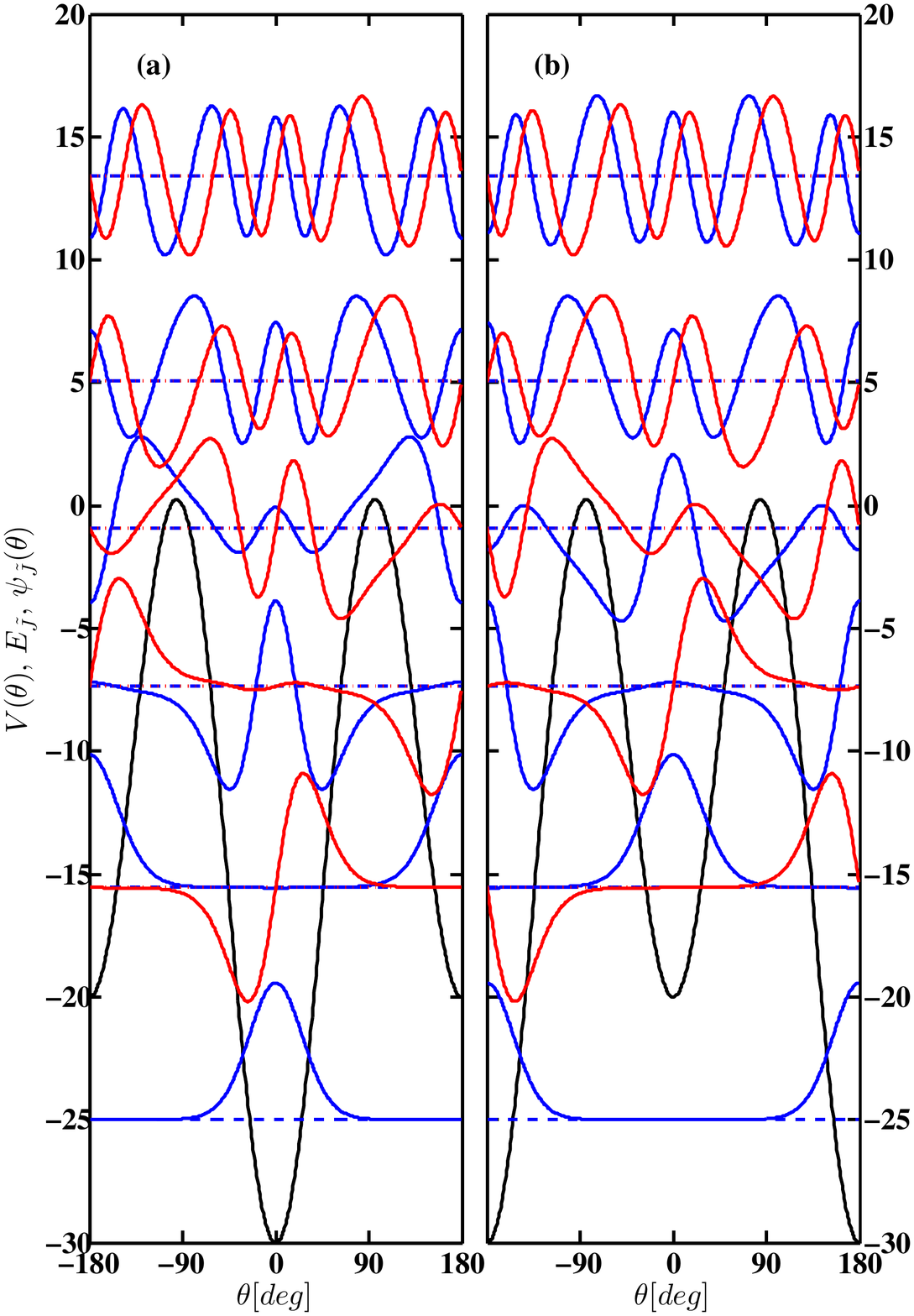}
	\caption{Superpartner potentials $V_1$ (left) and $V_2$ (right), eigenenergies (dashed lines) and eigenfunctions (full curves) of Hamiltonian (\ref{eq:hamilton}) with  $\eta=5,\zeta=25$, i.e., a first-order intersection with $\kappa=1$, see Sec. \ref{sec:kappa1}.}
	\label{fig:kappa1}
\end{figure}

The supersymmetric partner potentials obtained from Eq. (\ref{eq:eze}) are
\begin{eqnarray}
V_1(\theta) = - \beta\cos\theta - \beta^2\cos^2\theta \nonumber \\
V_2(\theta) = + \beta\cos\theta - \beta^2\cos^2\theta 
\label{eq:V12_1}
\end{eqnarray}
i.~e., $V_2(\theta)=V_1(\theta\pm\pi)$, as illustrated in Fig. \ref{fig:kappa1}. 
Hence, apart from a trivial interchange of the global and local minima, the partner potentials are identical and the ground state wavefunction of $H_2$ becomes
\begin{equation}
\psi_0^{(2)}(\theta) \propto \exp (-\beta\cos\theta)
\end{equation}
Although the SUSY partner Hamiltonians are completely isospectral, see also Tab. \ref{tab:ene}, one should not conclude that SUSY is broken \cite{s17} here, because the ground state wavefunctions $\psi_0^{(1)}(\theta)$ and $\psi_0^{(2)}(\theta)$ pertaining to both SUSY partner Hamiltonians $H_1$ and $H_2$ can nonetheless be annihilated by the intertwining operator $A$ and its adjoint $A^\dagger$:
\begin{equation}
A\psi_0^{(1)}=A^\dagger\psi_0^{(2)}=0
\label{eq:nihil_1}
\end{equation}
In contrast, there is a one-to-one pairing of all higher eigenstates, $n>0$, which we checked numerically
\begin{eqnarray}
A\psi_n^{(1)}\propto\psi_n^{(2)}\nonumber \\
A^\dagger\psi_n^{(2)}\propto\psi_n^{(1)}
\label{eq:twins_1}
\end{eqnarray}
where the odd parity of the intertwining operators $A$ and $A^\dagger$ implies a pairing of even eigenstates of $H_1$ with odd ones of $H_2$ and vice versa, see also Fig. \ref{fig:kappa1}.

The relation between $\eta$ and $\zeta$ established in Eq. (\ref{eq:eze_1}) implies that for {\it case A}, the topological index $\kappa=1$.
Hence the {\it case A} Hamiltonian gives rise to one single eigenstate while all its higher eigenstates occur as doublets. This we corroborated by a numerical solution to the Schr\"{o}dinger equation for the {\it case A} Hamiltonian, whose results for $(\eta,\zeta) = (5,25)$ are presented in Table \ref{tab:ene}.
As can also be seen in Fig. \ref{fig:kappa1}, the $n$-th doublet is comprised of a state with $n$ nodes near the global minimum ($\theta=0$) and a state with $n-1$ nodes near the local minimum ($\theta=\pi$), which is in agreement with eqs. (\ref{eq:loci}) and (\ref{eq:kappa}), thus rationalizing the occurrence of the first-order intersections characterized by $\kappa=1$. 

We note that the harmonic oscillator--like states centered at $\theta=0$ and at $\theta=\pi$ with quantum numbers $v_0$ and $v_\pi$ differing by one are of different parity. Hence, their coupling due the potentials $V_{1,2}$, which are of even parity, has to vanish, and hence these pairs of eigenstates are exactly degenerate. This contrasts with the case of a purely aligning interaction ($\kappa=0$), discussed in Sec. \ref{sec:zeta}, where we found a small but finite tunneling splitting.

\subsection {Second-order intersections or {\boldmath $\kappa=2$}}
\label{sec:kappa2}

For $\alpha=-1/2,\,\gamma=\pm1/2$ the superpotential (\ref{eq:W}) becomes
\begin{equation}
W^\pm=-\frac{1}{2}\cot\theta+\beta\sin\theta\pm\frac{1}{2}\csc\theta
\label{eq:W_2}
\end{equation}
and Eq. (\ref{eq:eze}) yields the following expressions for the interaction strength parameters and for the energy in terms of $\beta$,\begin{eqnarray}
\eta&=&2\beta \nonumber \\
\zeta&=&\beta^2 \nonumber \\
\epsilon^\pm &=& \frac{1}{4}\mp \beta-\beta^2
\label{eq:eze_2}
\end{eqnarray}
From Eq. (\ref{eq:psi_abc}) we obtain the {\it case B} eigenfunctions for the original potential $V_1$
\begin{eqnarray}
\psi_{\epsilon+}^{(1)}(\theta) &\propto& \cos\frac{\theta}{2}\exp (\beta\cos\theta) \nonumber \\
\psi_{\epsilon-}^{(1)}(\theta) &\propto& \sin\frac{\theta}{2}\exp (\beta\cos\theta)
\label{eq:psi_2}
\end{eqnarray}
Again, $\psi_{\epsilon+}^{(1)}$ is nodeless, corresponding to the ground state $\psi_0^{(1)}$ with energy $\epsilon^+=E_0^{(1)}$ which we will refer to as {\it case B.1}.
However, $\psi_{\epsilon-}^{(1)}$ exhibits a node at $\theta=0$ and hence corresponds to the first excited state $\psi_1^{(1)}$ with energy $\epsilon^-=E_1^{(1)}$ which we will refer to as {\it case B.2}.

The SUSY partner potentials for {\it cases B.1} and {\it B.2} take the form
\begin{eqnarray}
V_1^\pm(\theta)&=& -2\beta\cos\theta - \beta^2\cos^2\theta \nonumber   \\
V_2^+(\theta)&=& +\csc^2\theta -\cot\theta\csc\theta - \beta^2\cos^2\theta = \frac{1}{2}\sec^2\frac{\theta}{2}-\beta^2\cos^2\theta  \nonumber   \\
V_2^-(\theta)&=& +\csc^2\theta +\cot\theta\csc\theta - \beta^2\cos^2\theta = \frac{1}{2}\csc^2\frac{\theta}{2}-\beta^2\cos^2\theta
\label{eq:V12_2}
\end{eqnarray}
Note that $V_2^-(\theta)=V_2^+(\theta\pm\pi)$ and that the orientation field ($\propto\cos\theta$) is absent both in $V_2^+$ and in $V_2^-$.
In the vicinity of $\theta=0$, the former potential can be locally approximated by $1/2-\beta^2 \cos^2 \theta$ which is -- apart from an energy shift of 1/2 -- identical with our $\kappa=0$ case, i.e. the case of a pure alignment interaction discussed in Sec. \ref{sec:zeta}. 
This is also reflected by the numerical values shown in Tab. \ref{tab:ene}, i.~e., $-19.75 - 0.5 = -20.25$, which is quite close to -20.26 or -20.27 found numerically for $\kappa=0$.

\begin{figure}
	\centering
		\includegraphics[width=12cm]{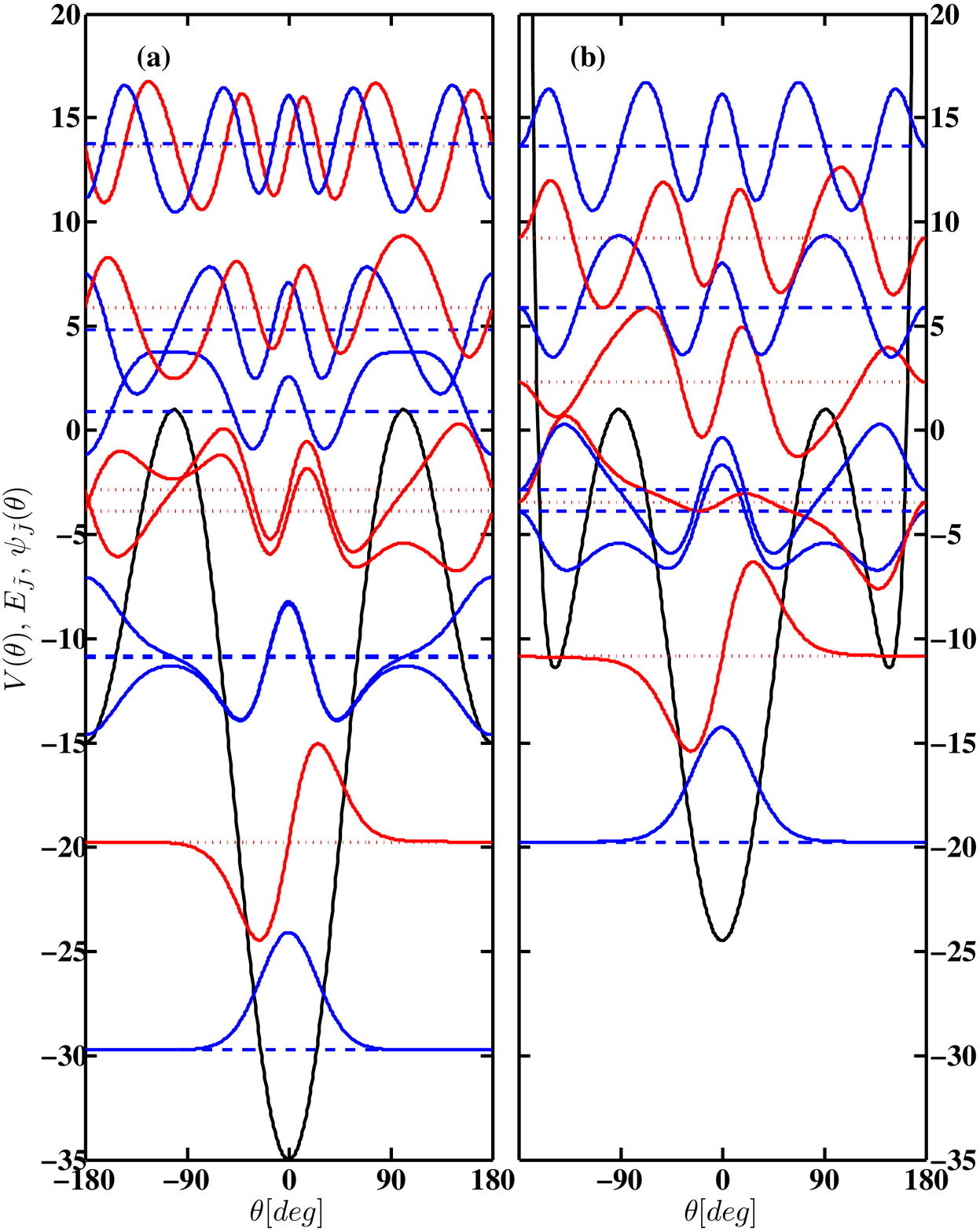}
	\caption{Superpartner potentials $V_1$ (left) and $V_2^+$ (right), eigenenergies (dashed lines) and eigenfunctions (full curves) of Hamiltonian (\ref{eq:hamilton}) with $\eta=10,\zeta=25$ for a second-order intersection, $\kappa=2$, see Sec. \ref{sec:kappa2}.}
	\label{fig:kappa2}
\end{figure}

Fig. \ref{fig:kappa2} shows the potentials $V_1$ and $V_2^+$ along with the corresponding wavefunctions. The ground and first excited states of $V_1$, see Eq. (\ref{eq:psi_2}), are found to be single states below the secondary minima. Both of them are annihiliated by the respective intertwining operators, $A^{\pm}\equiv d/d\theta + W^{\pm}$, pertaining to the superpotential $W^+$ for {\it case B.1} and $W^-$ for {\it case B.2},
\begin{eqnarray}
A^+\psi_0^{(1)}(\theta)=0 \nonumber \\
A^-\psi_1^{(1)}(\theta)=0
\end{eqnarray}
However, when $A^+$ acts on $\psi_1^{(1)}$, one obtains
\begin{equation}
A^+\psi_1^{(1)} \propto \psi_0^{(2)} \propto \sec\frac{\theta}{2}\exp (\beta\cos\theta)
\end{equation}
i. e., an analytic expression for the ground state wavefunction of the SUSY partner potential $V_2^+$ at energy $\epsilon^- = \frac{1}{4}+\beta-\beta^2$.
All higher states, $n>1$, of $V_1$ occur in nearly-degenerate tunneling pairs as long as they are bound (below the maxima of $V_1$). Like in the case of a purely aligning interaction (Sec. \ref{sec:zeta}), these states are followed by a single state (here at an energy of 0.9116) near the energetic barrier at $V_{\text{max}}=1$, whereas all unbound states form nearly-degenerate free-rotor-like pairs.
In general, for {\it case B} we observe that many but not all of the states of $V_1$ with $n>1$ have SUSY partner states of $V_2$ at the same energies and vice versa.  
Furthermore, there are no intertwining relations, such as Eq. (\ref{eq:twins_1}) in Sec. \ref{sec:kappa1}, for the wavefunctions any more. 
We note that, in general, the superpartner potentials (\ref{eq:V12_2}) are not expected to yield isospectral Hamiltonians because of the singularities in $W$ and $V_2$, which arise from the $\csc\theta$ term. However, it is known that in some such cases an accidental degeneracy between the spectra of $H_1$ and $H_2$ (at least partly) remains due to spatial symmetry, as explained in chapter 12 of ref. \cite{s17}.

The relation between $\eta$ and $\zeta$ established by Eq. (\ref{eq:eze_2}) implies that for {\it case B} the topological index $\kappa=2$.
Hence the {\it case B} Hamiltonian gives rise to two single eigenstates while all its higher eigenstates occur as doublets, see also the numerical data for $(\eta,\zeta) = (10,25)$ presented in Tab. \ref{tab:ene} and Fig. \ref{fig:kappa2}. 
It can also be gleaned from Fig. \ref{fig:kappa2} that the $n$--th doublet is comprised of a state with $n+1$ nodes near the global minimum ($\theta=0$) and another state with $n-1$ nodes near the local minimum ($\theta=\pi$), in agreement with eqs. (\ref{eq:loci}) and (\ref{eq:kappa}),  thereby rationalizing the occurrence of the second-order intersections characterized by $\kappa=2$. 
 In contrast to {\it case A} and the concomitant first-order intersections, we note that the harmonic oscillator--like states centered at $\theta=0$ and at $\theta=\pi$ with quantum numbers $v_0$ and $v_\pi$ differing by two are of same parity. Hence, their coupling induced by the (even-parity) potentials $V_{1,2}$ does not necessarily vanish, and the eigenstates occur in nearly degenerate pairs with a finite energy splitting.
However, these splittings converge to zero for increasing parameter $\beta$ as the harmonic librator limit is approached. 

\subsection {Third-order intersections or {\boldmath $\kappa=3$}}
\label{sec:kappa3}

For $\alpha=-1,\,\gamma=0$ the superpotential (\ref{eq:W}) becomes
\begin{equation}
W=-\cot\theta+\beta\sin\theta
\label{eq:W_3}
\end{equation}
and so Eq. (\ref{eq:eze}) yields the following expressions for the interaction parameters and the energy in terms of the coefficient $\beta$
\begin{eqnarray}
\eta&=&3\beta \nonumber \\
\zeta&=&\beta^2 \nonumber \\
\epsilon &=& 1-\beta^2
\label{eq:eze_3}
\end{eqnarray}
Upon substituting from Eq. (\ref{eq:eze_3}) into Eq. (\ref{eq:psi_abc}), we obtain the {\it case C} eigenfunction (corresponding to energy $\epsilon$) of the original Hamiltonian $H_1$
\begin{equation}
\psi_\epsilon^{(1)}(\theta) \propto \sin\theta\exp (\beta\cos\theta)
\label{eq:psi_3}
\end{equation}
which has a node at $\theta=0$, i.e., pertains to the first excited state $\psi_\epsilon^{(1)}=\psi_1^{(1)}$ with energy $\epsilon=E_1^{(1)}$. 

\begin{figure}
	\centering
		\includegraphics[width=12cm]{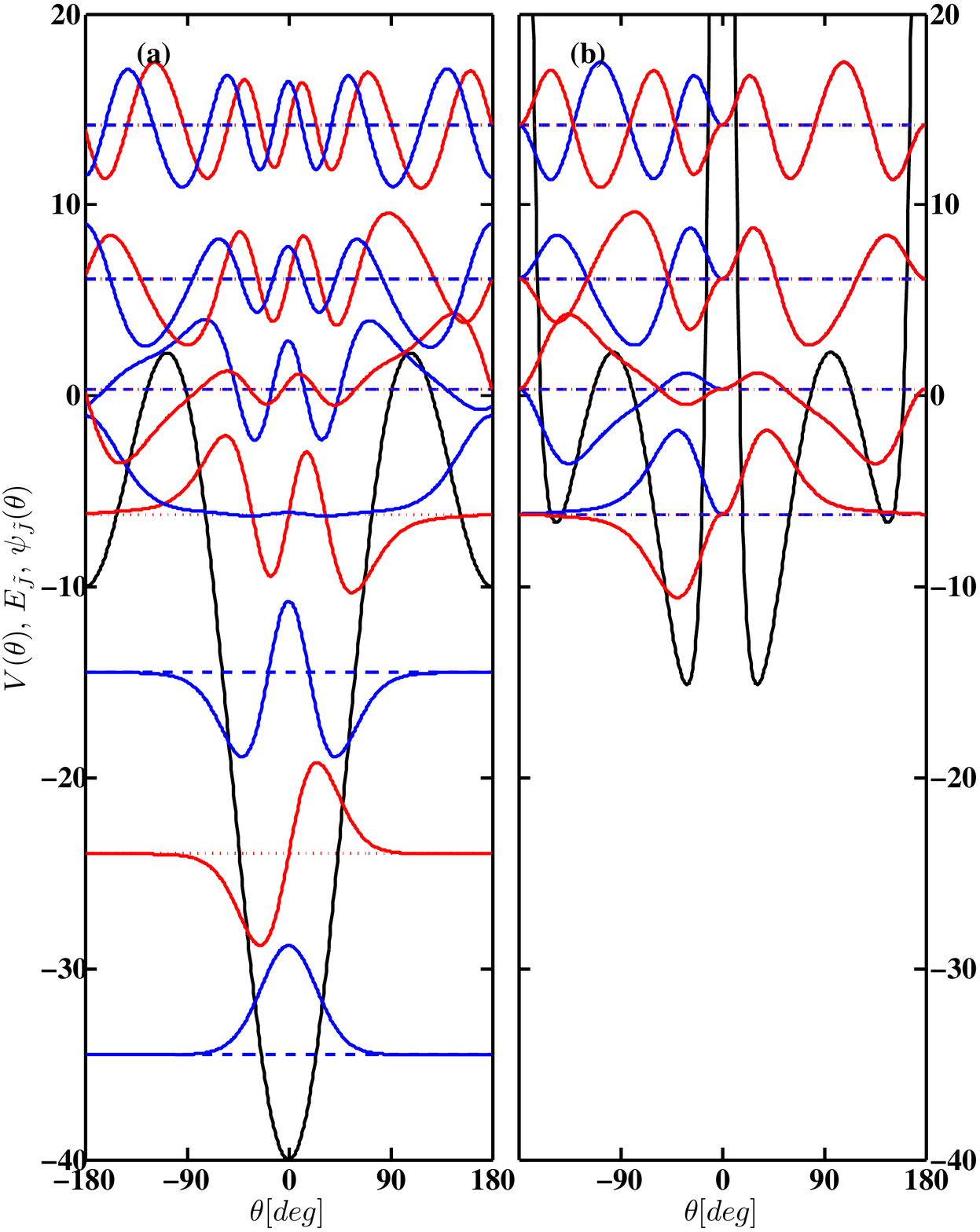}
	\caption{Superpartner potentials $V_1$ (left) and $V_2$ (right), eigenenergies (dashed lines) and eigenfunctions (full curves) of Hamiltonian (\ref{eq:hamilton}) with $\eta=15,\zeta=25$ for a third-order intersection, $\kappa=3$, see Sec. \ref{sec:kappa3}.}
	\label{fig:kappa3}
\end{figure}

The corresponding supersymmetric partner potentials for {\it case C} take the form
\begin{eqnarray}
V_1(\theta) =&               &-3\beta\cos\theta-\beta^2\cos^2\theta  \nonumber \\
V_2(\theta) =& 2\csc^2\theta & -\beta\cos\theta  -\beta^2\cos^2\theta
\label{eq:V12_3}
\end{eqnarray}
i.~e., the prefactor of the orientation field ($\propto\cos\theta$) is effectively reduced from $3\beta$ in $V_1$ to $\beta$ in $V_2$. 
The left panel of Fig. \ref{fig:kappa3} displays the large difference, of $6\beta$, in the well depths at the  global and local minima of the original potential $V_1$.
It accommodates three single states below the local minima which cannot have SUSY partner states because the wells of $V_2$ are much too shallow; however, only the middle one of the three states is annihilated by the intertwining operator
\begin{equation}
A\psi_1^{(1)}=0
\label{eq:nihil_3}
\end{equation}
All higher states, $n>2$, of $V_1$ occur in degenerate pairs with SUSY partner states for $V_2$ at exactly the same energies.  
However, there are no intertwining relations for their wavefunctions. 
As we have already noted in Sec. \ref{sec:kappa2}, the superpartner potentials (\ref{eq:V12_3}) do not necessarily yield isospectral Hamiltonians because of the singularities in $W$ and $V_2$.  At any rate, we find that in {\it case C}, $H_1$ and $H_2$ are isospectral, except for the lowest three states which are absent for $H_2$. 

The relation between $\eta$ and $\zeta$ established by Eq. (\ref{eq:eze_3}) implies that for {\it case C} the topological index $\kappa=3$.
Hence the {\it case C} Hamiltonian gives rise to three single eigenstates while all its higher eigenstates occur as doublets, see also the numerical data for $(\eta,\zeta) = (15,25)$ presented in Tab. \ref{tab:ene}. 
As can be also seen in Fig. \ref{fig:kappa3}, the $n$-th doublet arises from a state with $n+2$ nodes near the global minimum ($\theta=0$) and another state with $n-1$ nodes near the local minimum ($\theta=\pi$). This is in agreement with eqs. (\ref{eq:loci}) and (\ref{eq:kappa}), thus rationalizing the occurrence of the third-order intersections, $\kappa=3$. 
We note that the harmonic oscillator--like states centered at $\theta=0$ and $\theta=\pi$ with quantum numbers $v_0$ and $v_\pi$ differing by three are of different parity. Hence, their coupling induced by the (even parity) potentials $V_{1,2}$ has to vanish, and the eigenstates occur in exactly degenerate pairs without a tunneling splitting.

A special case of {\it case C} is the free rotor, which arises for $\beta=0$. Although its analytic eigenenergies, $E_n^{(1)}=n^2$, and eigenfunctions $\propto\sin n\theta, \propto\cos n\theta$ are well-known, it is nevertheless instructive to discuss the free rotor case from the SUSY point of view, see also ref. \cite{s17}. 

The {\it case C} superpotential (\ref{eq:W_3}) reduces to 
\begin{equation}
W_1=-\cot\theta
\label{eq:W_3free1}
\end{equation}
and the wavefunction (\ref{eq:psi_3}) simplifies to
\begin{equation}
\psi_1^{(1)}(\theta) \propto \sin\theta
\label{eq:psi_3free1}
\end{equation}
with a corresponding energy $E_1^{(1)}=1$. According to Eq. (\ref{eq:V12_3}), the vanishing potential, $V_1=0$, of the free rotor has the following SUSY partner 
\begin{equation}
V_2(\theta) = 2\csc^2\theta  
\label{eq:V2_3free1}
\end{equation}
which has bound states at energies $E_n^{(2)}=(n+2)^2$, i.e., again the three lowest states of the original (free rotor) potential with eigenenergies $E_{0,1,2}^{(1)}=0,1,1$ lie below the minimum, $V_{\rm min}^{(2)}=2$, of the partner potential which, therefore, cannot have SUSY partner states. 

The SUSY procedure can be repeated, i.e., one can find a new superpotential such that $V_2=W_2^2-W'_2+\epsilon$ which yields
\begin{equation}
W_2=-2\cot\theta
\label{eq:W_3free2}
\end{equation}
whose partner potential, $V_3=W_2^2+W_2'+\epsilon$, evaluates to
\begin{equation}
V_3(\theta) = 6\csc^2\theta  
\label{eq:V2_3free2}
\end{equation} 

This procedure can be repeated to yield   the $n$-th potential
\begin{equation}
V_n(\theta) = n(n-1)\csc^2\theta  
\label{eq:Vnfree}
\end{equation} 
and the $n$-th superpotential
\begin{equation}
W_n(\theta) = n\cot\theta  
\label{eq:Wnfree}
\end{equation} 
for the free-rotor case.
The eigenenergy
\begin{equation}
\epsilon_n = n^2
\label{eq:Enfree}
\end{equation} 
pertains to the ground state for states with $n \ge 2$ and to the first excited state for $n=1$. As a result, the corresponding Hamiltonians $H_n$ are isospectral, except that, starting with $H_2$, each subsequent Hamiltonian has one level (two bound states) less than the previous one. For $H_1$, as many as one state with $n=0$ and two states with $n=1$ are abandoned, i.e., $H_2$ has two levels (three bound states) less that $H_1$. 

We note that the free-rotor superpotential (\ref{eq:Wnfree}) is closely related to the Rosen-Morse I superpotential  of ref. \cite{s17}, which is likewise shape invariant. This accounts for the exact solvability of the free-rotor problem.

\section{Conclusions and Outlook}
\label{outlook}

We undertook mutually complementary  analytic and computational study of the planar quantum rotor subject to combined orienting and aligning interactions, characterized, respectively, by dimensionless parameters $\eta$ and $\zeta$. We considered a full range of interaction strengths, which convert, jointly or separately, the planar rotor into a planar hindered rotor or a planar quantum pendulum or a planar harmonic librator (angular harmonic oscillator), depending on the values of $\eta$ and $\zeta$. Following upon our previous study of the corresponding problem in 3D (spherical rotor/pendulum), we were concerned with the topology of the eigenenergy surfaces spanned by the interaction parameters $\eta$ and $\zeta$ as well as with the supersymmetry of the planar eigenproblem as a means for identifying its analytic solutions.

{\it Topology}. We found that the loci of all the intersections that arise among the eigenenergy surfaces of the planar quantum pendulum are accurately rendered by a simple formula, Eq. (\ref{eq:loci}). The formula is accurate despite the fact that its derivation was based on two rather crude approximations, namely a harmonic approximation of the combined orienting and aligning potentials and the approximate equality of the vibrational quanta in eqs. (\ref{eq:E_harm_0}) and (\ref{eq:E_harm_pi}) for $2\zeta\gg\eta$. Furthermore, since the equation for the loci, Eq. (\ref{eq:loci}), and the definition, Eq. (\ref{eq:kappa}),  of the topological index $\kappa$ are independent of the eigenstate, the energy levels exhibit a general pattern that only depends on the values of $\kappa$: for each $\kappa$, there are $\kappa$ single states, followed, in ascending order, by all other states which are doubly degenerate. This energy level pattern reflects the fact that above the local minimum, states can be bound by both the local ($\frac{\pi}{2}\le \theta \le \pi$) and global minima ($\theta = 0$) whereas below the local minimum states can only be bound by the global minimum. Since
the energy difference between the global and local minima increases linearly with $\kappa$, the number of single states bound solely by the global minimum increases with $\kappa$ as well (in fact is equal to $\kappa$). States bound by both the global and local minima that lie below the maximum of the potential, Eq. (\ref{eq:Vmax}), occur as doublets. Interestingly, the above eigenenergy level pattern persists even for such values of $\kappa\approx 10$ ($2\zeta<\eta$) where no local minima occur. And finally, the intersections are found to be
genuine for odd $\kappa$ and avoided for even $\kappa$. This is due to the fact that for even $\kappa$, the intersecting levels are of same parity and thus can be coupled by the combined interactions potential which is of even parity.  For odd $\kappa$, the intersecting states are of opposite parity and so cannot be coupled by the even-parity potential.

{\it Supersymmetry.} By invoking supersymmetric quantum mechanics (SUSY QM), we have identified three sets of conditions ({\it cases A, B} and {\it C}) under which the eigenproblem for Hamiltonian (\ref{eq:hamilton}) pertaining to the planar quantum pendulum can be solved analytically. As it turns out,  each of the {\it cases} implies a certain ratio of the interaction parameters $\eta$ and $\zeta$ and, thereby, a certain value of the topological index $\kappa$. This made it possible to identify each case with a particular topology: {\it case A} with $\kappa=1$, {\it case B} with $\kappa=2$, and {\it case C} with $\kappa=3$.  Whereas {\it cases A} and {\it B.1} furnish the ground-state wavefunctions, {\it cases B.2} and {\it C} furnish the first excited-state wavefunctions. The free planar rotor has been identified as a subcase of {\it case C}, one which exhibits shape invariance and therefore analytic solvability for all states. 

By making use of the analytic wave functions, we evaluated, likewise in analytic form, the observables of interest, such as the expectation values of the angular momentum squared, the orientation and alignment cosines, and the corresponding eigenenergy. These are summarized in Table \ref{table:parameters}.
By virtue of the SUSY QM apparatus, we constructed for each potential $V_1$ its supersymmetric partner potential $V_2$ and for {\it cases A} and {\it B.1} found the $V_2$ ground-state wave functions in analytic form. 
Apart from the singularities introduced by the terms proportional to $\csc \theta$, the main difference between $V_1$ and $V_2$ is that the orienting field ($\eta$) is effectively reduced by $2 \beta$, which tends to decrease the well depth of the global minimum while increasing the well depth of the local/secondary minimum. This reduced asymmetry of $V_2$ compared with $V_1$  is the deeper reason why (a certain number of) single states (which are always localized around the global minimum) present in $V_1$ are absent in $V_2$. However, while in standard SUSY QM \cite{s17,s26} only one state is eliminated upon the transition from $V_1$ to $V_2$, our present analysis shows that the planar pendulum problem somewhat deviates from this pattern.
In our {\it case A} ($\kappa=1$), we find strictly isospectral (no state eliminated) partner Hamiltonians although SUSY is not broken.
In {\it case B} ($\kappa=2$), there is one state eliminated, as expected, but a one-on-one correspondence between higher eigenstates of $V_1$ and $V_2$ is incomplete.  
In {\it case C} ($\kappa=3$), we find that the transition to $V_2$ eliminates no less than the lowest three states! We came across a similar pattern when inspecting Fig. 12.1 of ref. \cite{s17} for the harmonic oscillator, where it is loosely attributed to the spatial symmetry of the problem. 
It is therefore likely that our findings about the planar quantum pendulum are related to the symmetry (parity) of the problem and the resulting degeneracy patterns as well. Clearly, more work needs to be done here, with the ultimate goal of developing a theory that combines  supersymmetry and spatial symmetry, $SUSY+SY=SUSYSY$.

Last but not least, we note that it has not escaped our notice  that our {\it cases A}, {\it B}, and {\it C} are equivalent to the spherical pendulum problem studied in ref. \cite{LemMusKaisFriNJP2011} for $m=-\frac{1}{2}$, $m=0$, and $m=\frac{1}{2}$, respectively. However, despite this similarity, there is an important difference: in the 3D, spherical case, the polar angle is only defined on a half-circle ($0 \le \theta \le \pi$), with repercussions for  symmetry (e.g., all crossings of the eigenenergy surfaces in 3D are avoided). In our forthcoming paper, we revisit the spherical pendulum case.

\begin{acknowledgments}
Support by the \textit{Deutsche Forschungsgemeinschaft} (DFG) through grants SCHM 1202/3-1 and FR 3319/3-1 is gratefully acknowledged.
\end{acknowledgments}

\bibliography{CombFields}

\clearpage

\begin{turnpage}
\begin{table}
	\centering
		\begin{tabular}{cccccc}
		  \hline \hline
		  Eigenvalue & Function & Transformation       & period & parity & parity    \\
		  $r=0,1,\ldots$  & $m=0,1,\ldots$  &        &        & ($x=0$)& ($x=\pi/2$)\\
		  \hline
			$a_{2r}$   & $ce_{2r  }(x;q)=\sum A^{(2r  )}_{2m  }(q)\cos 2mx$     & $ce_{2r}  (x;-q) = (-1)^r ce_{2r}  (\pi/2-x;q)$ & $\pi$  & even & even  \\
			$b_{2r+1}$ & $se_{2r+1}(x;q)=\sum B^{(2r+1)}_{2m+1}(q)\sin (2m+1)x$ & $se_{2r+1}(x;-q) = (-1)^r ce_{2r+1}(\pi/2-x;q)$ & $2\pi$ & even & odd   \\
			$a_{2r+1}$ & $ce_{2r+1}(x;q)=\sum A^{(2r+1)}_{2m+1}(q)\cos (2m+1)x$ & $ce_{2r+1}(x;-q) = (-1)^r se_{2r+1}(\pi/2-x;q)$ & $2\pi$ & odd  & even  \\
			$b_{2r+2}$ & $se_{2r+2}(x;q)=\sum B^{(2r+2)}_{2m+2}(q)\sin (2m+2)x$ & $se_{2r+2}(x;-q) = (-1)^r se _{2r+2}(\pi/2-x;q)$ & $\pi$  & odd  & odd   \\
		  \hline \hline
		\end{tabular}
	\caption{Mathieu cosine elliptic ($ce$) and sine elliptic ($se$) functions. The third column shows the relations between functions for positive and negative $q$. The last three columns list the symmetry properties that pertain to the case of negative $q$. The four classes of Mathieu functions are ordered according to their ascending eigenvalues (energetic ordering). Adapted from ref. \cite{Gutierrez:03a}.}
	\label{tab:mathieu}
\end{table}
\end{turnpage}

\begin{table}
\begin{center}
\begin{tabular}{cdddd}
\hline 
\hline
$n$ & \multicolumn{1}{c}{$\kappa=0$} & \multicolumn{1}{c}{$\kappa=1$} & \multicolumn{1}{c}{$\kappa=2$} & \multicolumn{1}{c}{$\kappa=3$} \\
\hline

0 & -20.2670 &\textbf{-25} & \textbf{-29}.\textbf{75} & -34.5125   \\
1 & -20.2629 &-15.5485     & \textbf{-19}.\textbf{75} & \textbf{-24} \\
2 & -11.4689 &-15.5485     & -10.8997        & -14.4875     \\
3 & -11.3496 & -7.3631     & -10.8118        &  -6.1992     \\
4 &  -4.5570 & -7.3631     &  -3.8735        &  -6.1992     \\
5 &  -3.3987 & -0.9486     &  -2.8667        &   0.3568     \\
6 &   0.1931 & -0.9486     &   0.9116        &   0.3568     \\
7 &   4.3920 &  5.0669     &   4.8582        &   6.1324     \\
8 &   5.3588 &  5.0669     &   5.9178        &   6.1324     \\
9 &  13.2809 & 13.4317     &  13.6472        &  14.1848     \\
10 & 13.3948 & 13.4317     &  13.7747        &  14.1848     \\
\hline \hline
	\end{tabular}
\end{center}

\caption{\small Eigenenergies for the Hamiltonian of Eq. (\ref{eq:hamilton}) with $\zeta=\beta^2=25$ and  $\eta=\kappa \beta$, as also displayed in Figs. \ref{fig:kappa0} (right panel) and in Figs. \ref{fig:kappa1}-\ref{fig:kappa3}. Calculated with the Fourier Grid Hamiltonian (FGH) method \cite{Meyer:70a,Marston:89a} as implemented in WavePacket software \cite{Schmidt-WavePacket4.9} with 512 grid points. Energies above 10000 have been truncated. Cases where analytic eigenenergies/wavefunctions are available are printed in bold face, see also equations (\ref{eq:eze_1}), (\ref{eq:eze_2}), and (\ref{eq:eze_3}). For $\kappa=0$, the $n=0,1,2,3,...$ states corresponds to Mathieu's states $ce_0$, $se_1$, $ce_1$, $se_2$, ...  for a purely aligning interaction. We note thast no other analytic solutions were found for value of $\kappa \le 10$ and $n \le 20$.}
\label{tab:ene}
\end{table}

\begin{turnpage}
\begin{table}
\centering
\begin{tabular}{| c | c | c | c | c | c | c | c | c |}
\hline 
\hline
& $\kappa=1$ & $\kappa=2$  & $\kappa=2$  & $\kappa=3$  \\[5pt]
\hline

$W$ & $\beta \sin\theta$ &  $\frac{1}{2}\cot\theta+\beta\sin\theta+\frac{1}{2}\csc\theta$ & $\frac{1}{2}\cot\theta+\beta\sin\theta-\frac{1}{2}\csc\theta$ & $-\cot\theta+\beta\sin\theta$  \\[5pt]

$\eta$ & $\beta$  &  $2\beta$ & $2\beta$ & $3\beta$  \\[5pt]

$\zeta$ & $\beta^2$  &  $\beta^2$ & $\beta^2$ & $\beta^2$  \\[5pt]

$\psi$ & $\psi_0\propto e^{\beta  \cos (\theta )}$ & $\psi_0\propto \cos \left(\frac{\theta }{2}\right) e^{\beta  \cos (\theta )}$ & $\psi_1\propto \sin \left(\frac{\theta }{2}\right) e^{\beta  \cos (\theta )}$ & $\psi_1\propto \sin (\theta ) e^{\beta  \cos (\theta )}$  \\[5pt]

$N$ & $2 \pi  I_0(2 \beta )$ & $\pi  (I_0(2 \beta )+I_1(2 \beta ))$ & $\pi  (I_0(2 \beta )-I_1(2 \beta ))$ & $\pi  \, _0\tilde{F}_1\left(;2;\beta ^2\right)$  \\[5pt]

$\langle \cos\theta \rangle$ &  $\frac{I_1(2 \beta )}{I_0(2 \beta )}$  &  $\frac{\pi  I_2(2 \beta )+\frac{1}{2} \pi  (2 \beta +1) \, _0\tilde{F}_1\left(;2;\beta ^2\right)}{\pi  (I_0(2 \beta )+I_1(2 \beta ))}$  & $\frac{\frac{1}{2} \pi  (2 \beta -1) \, _0\tilde{F}_1\left(;2;\beta ^2\right)-\pi  I_2(2 \beta )}{\pi  (I_0(2 \beta )-I_1(2 \beta ))}$ & $\frac{I_2(2 \beta )}{\beta  \, _0\tilde{F}_1\left(;2;\beta ^2\right)}$ \\[5pt]

$\langle \cos^2\theta \rangle$ &  $\frac{2 I_2(2 \beta )+\, _0\tilde{F}_1\left(;2;\beta ^2\right)}{2 I_0(2 \beta )}$ &  $\frac{(2 \beta +1) \, _0\tilde{F}_1\left(;2;\beta ^2\right)+\beta  (2 \beta -1) \, _0\tilde{F}_1\left(;3;\beta ^2\right)}{2 (I_0(2 \beta )+I_1(2 \beta ))}$ & $\frac{(1-2 \beta ) I_1(2 \beta )+(2 \beta +1) I_2(2 \beta )}{2 \beta  (I_0(2 \beta )-I_1(2 \beta ))}$ & $\frac{\pi  \, _0\tilde{F}_1\left(;2;\beta ^2\right)-\frac{3 \pi  \, _0\tilde{F}_1\left(;3;\beta ^2\right)}{2}}{\pi  \, _0\tilde{F}_1\left(;2;\beta ^2\right)}$ \\[5pt]

$\langle {\bf{J}}^2 \rangle$ &  $\frac{\beta  I_1(2 \beta )}{2 I_0(2 \beta )}$  &  $\frac{(2 \beta +1) I_0(2 \beta )+(2 \beta -1) I_1(2 \beta )}{4 (I_0(2 \beta )+I_1(2 \beta ))}$ & $\frac{(1-2 \beta ) I_0(2 \beta )+(2 \beta +1) I_1(2 \beta )}{4 (I_0(2 \beta )-I_1(2 \beta ))}$ & $\frac{\frac{3}{2} \pi  I_2(2 \beta )+\pi  \, _0\tilde{F}_1\left(;2;\beta ^2\right)}{\pi  \, _0\tilde{F}_1\left(;2;\beta ^2\right)}$ \\[5pt]

$\epsilon$ &  $E_0=-\beta^2$ &  $E_0=-\beta ^2-\beta +\frac{1}{4}$ & $E_1=-\beta ^2+\beta +\frac{1}{4}$ & $E_1=-\beta^2+1$  \\[5pt]

\hline
\hline
\end{tabular}
\caption{\small Superpotentials, $W$, interaction parameters, $\eta$ and $\zeta$, wave functions, $\psi$, normalization factors, $N$, orientation cosines, $\langle\cos\theta\rangle$, alignment cosines $\langle\cos\theta\rangle$, the expectation values of the angular momentum squared, $\langle {\bf{J}}^2 \rangle$, and the eigenenergies, $\epsilon$, for the the four cases that yield an analytic solution to the eigenproblem of Hamiltonian (1). $I_m(z)$ are the modified Bessel functions and  $_0\tilde{F}_1\left(;2;\beta ^2\right)$ are the confluent hypergeometric functions.}
\label{table:parameters}
\end{table}
\end{turnpage}

\clearpage

\end{document}